\documentclass[12pt]{article}\usepackage{qInfoStyle}

\renewcommand{\BeginSection}[1]{
	\newpage
	\section{#1}
	\setcounter{equation}{0}
	\setcounter{figure}{0}
}
\renewcommand{\EndSection}[0]{
}

\begin{document}


\title{Quantum Information Theory - \\
A Quantum Bayesian Net Perspective}

\author{Robert R. Tucci\\
        P.O. Box 226\\ 
        Bedford,  MA   01730\\
        tucci@ar-tiste.com}

\date{ \today} 

\maketitle

\vskip2cm
\section*{Abstract}
The main goal of this paper is to give a pedagogical introduction
to Quantum Information Theory---to do this in a new way, using network diagrams called
Quantum Bayesian  Nets. A lesser goal of the paper
is to propose a few new ideas, such as
associating with each quantum Bayesian net 
a very useful density matrix that we call the meta density matrix.

\newpage

\EndSection

\tableofcontents


\BeginSection{Introduction}\label{introduction}







The main goal of this paper is to give a pedagogical introduction
to Quantum Information Theory---to do this in a new way, using network diagrams called
Quantum Bayesian (QB) Nets. The paper assumes no prior knowledge of 
Classical\cite{Mans}-\cite{Cover} or Quantum\cite{Hel}-\cite{SchuLecs}
Information Theory.  It does assume a 
good understanding of the machinery of Quantum Mechanics,
such as one would obtain by reading any reasonable textbook that 
explains Dirac bra-ket formalism. The paper reviews QB nets in an appendix.
If you have difficulty understanding said appendix,
you might want to read Ref.\cite{Tucci95} before continuing this paper.

Most of the ideas discussed in this paper are not new. They are
well-known, standard ideas invented by the pioneers (Bennett, Holevo,
Peres, Schumacher, Wootters, etc.) of the field of Quantum Information Theory.
What is new about this paper is that, whenever possible and advantageous, we rephrase 
those ideas in the visual language of QB nets. 
The paper does present a few new ideas, such as
associating with each QB net a very useful density matrix that we call the meta density matrix of the net.

The topics covered in this paper are shown in the Table of Contents. 
The paper, in its present form,  
is far from being a complete account of the field
of Quantum Information Theory. Some important topics that were left out (because the author
didn't have enough time to write them up) are:
quantum compression, quantum error correction, channel capacities, quantum approximate cloning,
entanglement quantification and manipulation. 
Future editions of this paper may include some of these topics. 
I welcome any suggestions or comments.
To fill in gaps left by this paper, or to find alternative explanations of difficult topics, see 
Refs.\cite{Hel}-\cite{SchuLecs} and references therein.

\EndSection


\BeginSection{Notation}\label{notation}

In this section, we will introduce certain notation which is 
used throughout the paper. 

We define $Z_{a,b}$ to be the set  $\{a, a+1, \cdots, b\}$ for 
any integers $a$ and $b$. Let $Bool = \{0,1\}$. For any finite set $S$, let $|S|$ denote the number of elements
in $S$.

The 
Kronecker delta function $\delta(x,y)$ equals one if $x=y$ and zero otherwise.
We will often abbreviate $\delta(x,y)$ by $\delta^x_y$.

We will often use the symbol 
$\sum_{ri}$ to mean that one must sum
whatever is on the right-hand side of this symbol 
over all repeated indices (a sort of Einstein summation convention).
Likewise, $\sum_{all}$ will mean that one should sum over all indices.
If we wish to exclude a particular index from the summation,
we will indicate this by a slash followed by the name of the index.
For example, in $\sum_{ri/f}$ or $\sum_{all/f}$ we wish to exclude summation over $f$.

The Pauli matrices $\sigma_x$, $\sigma_y$ and $\sigma_z$ are defined by

\beq
\sx=
\left(
\begin{array}{cc}
0&1\\
1&0
\end{array}
\right)
\;,
\;\;
\sy=
\left(
\begin{array}{cc}
0&-i\\
i&0
\end{array}
\right)
\;,
\;\;
\sz=
\left(
\begin{array}{cc}
1&0\\
0&-1
\end{array}
\right)
\;.
\eeq

For any real $p\in [0,1]$, we define the binary entropy function $h(\cdot)$ by

\beq
h(p) = -p\log_2(p) - (1-p)\log_2 (1-p)
\;.
\eeq

When speaking of bits with states
0 and 1, we will often use an overbar to represent the opposite state: 
$\bar{0} = 1$, $\bar{1} = 0$.

We will underline random variables. For example,
we might write $P(\rva = a)$ for the probability that
the random variable $\rva$ assumes value $a$.
$P(\rva=a)$ will often be abbreviated by $P(a)$ when no 
confusion will arise. $S_\rva$ will denote the
set of values which the random variable $\rva$
may assume, and $N_\rva$ will denote the number of
elements in $S_\rva$. With each random variable $\rva$,
we will associate an orthonormal basis $\{ \ket{a} | a\in S_\rva \}$
which we will call the {\it $\rva$ basis}. We will 
represent by $\hil_\rva$ the Hilbert space spanned by the $\rva$ basis.
$\ket{\rva=a}$ will mean the same thing as $\ket{a}$; 
$\ket{\rva=a}$ is just a more explicit notation that indicates that $\ket{a}$ belongs to $\hil_\rva$.
If $\rvx_1, \rvx_2, \cdots, \rvx_N$ are any $N$ random variables,
we will use $\hil_{\rvx_1, \rvx_2, \cdots, \rvx_N}$
to denote $\hil_{\rvx_1}\otimes \hil_{\rvx_2}\otimes \cdots \hil_{\rvx_N}$.

Whenever we use the word ``ditto", as in ``X (ditto, Y)", we mean
that the statement is true if X is replaced by Y. For example,
if we say ``A (ditto, X) is smaller than B (ditto, Y)", we mean 
``A is smaller than B" and ``X is smaller than Y".

This paper will also utilize certain notation
associated with classical and quantum Bayesian nets.
See Appendix \ref{app:bnet-review} for a review of
such notation.

\EndSection


\BeginSection{Classical Entropy: Its Definition and Properties}\label{sec:c-ent}

In this section, we will define various classical entropies  associated with
a CB net.

Suppose $p_1, p_2, \ldots, p_N$  are $N$ non-negative numbers which add up to one. 
The {\it classical entropy} $H(\vec{p})$ of $\vec{p}$ is defined by

\beq
H(\vec{p}) = -\sum_{i=1}^{N} p_i \log_2 p_i
\;.
\label{eq:CE-h-def}\eeq
$H(\vec{p})$ measures the spread  of the probability distribution $\vec{p}$. 

In Thermodynamics, entropy measures the disorder of a macroscopic system. See
Ref.\cite{PeresBook} for a discussion of the relationship between the entropy 
of Thermodynamics and Eq.(\ref{eq:CE-h-def}).

In Communication Theory, one uses the words
``information" and ``entropy" interchangeably.
In the context of communication theory,
the word ``information" means {\it information content of an average message}.
Given any random variable $\rvx$, one may think of
a sequence $x_1, x_2, \ldots, x_N$ of samples 
of $\rvx$ as a message. Then one makes the assumption that 
the more information an average message (of fixed length) carries, the higher the variance of $\rvx$ will be,
and vice versa. Eq.(\ref{eq:CE-h-def}) quantifies the variance of $\rvx$ if 
we replace $p_i$ and the sum over $1\leq i \leq N$ by  $P(\rvx = x)$ and a sum over $x\in S$,
where $S$ is the set of values that $\rvx$ can assume.

When dealing with a CB net, it is convenient to rephrase 
Eq.(\ref{eq:CE-h-def}) in terms of the node random variables of the net. 
Consider a CB net $\cnet$  with $N$ nodes labelled by the
random variables $\rvx_1, \rvx_2, \dots, \rvx_N$.
These $N$ random variables are related by a joint probability distribution $P(x.)$.
Suppose $\Gamma_1$ and $\Gamma_2$ are  non-empty subsets of $\zn$. 
$\Gamma_1$ and $\Gamma_2$ need not be disjoint.
The probability distributions $P[ (x.)_{\Gamma_1}]$,  $P[(x.)_{\Gamma_2}]$ 
and $P[(x.)_{\Gamma_1 \cup \Gamma_2}]$
can be obtained by summing $P(x.)$ over the unwanted arguments,
a process called marginalization.
We define:

\beq
H[(\rvx.)_{\Gamma_1}] = -\sum_{(x.)_{\Gamma_1}} P[(x.)_{\Gamma_1}] \log_2 P[(x.)_{\Gamma_1}] 
\;,
\label{eq:CE-h-simple}\eeq

\beq
H[(\rvx.)_{\Gamma_1}  |  (\rvx.)_{\Gamma_2}] =
 -\sum_{(x.)_{\Gamma_1 \cup \Gamma_2}} P[(x.)_{\Gamma_1 \cup \Gamma_2}] 
 \log_2 \left(\frac{P[(x.)_{\Gamma_1 \cup \Gamma_2}]}{P[(x.)_{\Gamma_2}]}\right)
\;,
\label{eq:CE-h-cond}\eeq

\beq
H[(\rvx.)_{\Gamma_1} : (\rvx.)_{\Gamma_2}] = \sum_{(x.)_{\Gamma_1 \cup \Gamma_2}} P[(x.)_{\Gamma_1 \cup \Gamma_2}]
 \log_2 \left(\frac{P[(x.)_{\Gamma_1 \cup \Gamma_2}]}{P[(x.)_{\Gamma_1}]P[(x.)_{\Gamma_2}]}\right)
\;.
\label{eq:CE-h-mutual}\eeq
For example, if $\rva$ and $\rvb$ are nodes of a CB net, then
\beq
H(\rva) = -\sum_a P(a) \log_2 P(a)
\;,
\eeq

\beq
H(\rva, \rvb) = -\sum_{a,b} P(a,b) \log_2 P(a,b)
\;,
\eeq

\beq
H(\rva  |  \rvb) = -\sum_{a,b} P(a,b) \log_2 P(a  |  b)
\;,
\eeq

\beq
H(\rva:\rvb) = \sum_{a,b} P(a,b) \log_2 \left(\frac{P(a,b)}{P(a)P(b)}\right)
\;,
\eeq
where  $P(a)= \sum_b P(a,b)$, $P(b)= \sum_a P(a,b)$, and the sums over $a$ (ditto, $b$) 
range over all $a\in S_\rva$ (ditto, $b\in S_\rvb$). 

Note that definitions
Eqs.(\ref{eq:CE-h-simple}) to (\ref{eq:CE-h-mutual}) are independent of the order of the 
node random variables within $(\rvx.)_{\Gamma_1}$ and $(\rvx.)_{\Gamma_2}$.
For example, if $\rva, \rvb, \rvc$ are nodes of a CB net, then

\beq
H(\rva, \rvb, \rvc) = H(\rva, \rvc, \rvb),
\;\;
H[\rva| (\rvb, \rvc)] = H[\rva| (\rvc, \rvb)]
\;.
\eeq
It is convenient to extend definitions Eqs.(\ref{eq:CE-h-simple}) to (\ref{eq:CE-h-mutual})
in the following two ways. First, we will allow
$(\rvx.)_{\Gamma_1}$ (ditto, $(\rvx.)_{\Gamma_2}$)
to contain repeated random variables. If it does, then 
we will throw out any extra copies of a random variable.
For example, if $\rva, \rvb, \rvc$ are nodes of a CB net, then

\beq
H(\rva, \rva, \rvb, \rvc) = H(\rva, \rvb, \rvc),
\;\;
H[\rva| (\rvb, \rvb, \rvc)] = H[\rva| (\rvb, \rvc)]
\;.
\eeq
Second, we will allow
$(\rvx.)_{\Gamma_1}$ (ditto, $(\rvx.)_{\Gamma_2}$)
to contain internal parentheses. If it does, then 
we will ignore the internal  parentheses.
For example, if $\rva, \rvb, \rvc, \rvd$ are nodes of a CB net, then

\beq
H[(\rva, \rvb), \rvc] = H(\rva, \rvb, \rvc),
\;\;
H[\rva| ((\rvb, \rvc), \rvd)] = H[\rva| (\rvb, \rvc, \rvd)]
\;.
\eeq

Let $\rvX = (\rvx.)_{\Gamma_1}$ and $\rvY = (\rvx.)_{\Gamma_2}$.
$H(\rvX)$ measures the spread of the $P(X)$ distribution. $H(\rvX  |  \rvY)$
is called the {\it conditional entropy} of $\rvX$ given $\rvY$. 
$H(\rvX : \rvY)$ is called the {\it mutual entropy} of $\rvX$ and $\rvY$, and
it measures the dependency of $\rvX$ and $\rvY$: it is non-negative, and it
equals zero iff $\rvX, \rvY$ are independent 
random variables (i.e., $P(X, Y) = P(X) P(Y)$ for all $X\in S_\rvX$ and $Y\in S_\rvY$).

Note that Eqs.(\ref{eq:CE-h-simple}) to (\ref{eq:CE-h-mutual}) imply that 
 
\beq
H(\rvX  |  \rvY) = H(\rvX, \rvY) - H(\rvY)
\;,
\label{eq:CE-decomp-cond}\eeq

\beq
H(\rvX : \rvY) = H(\rvX) + H(\rvY) - H(\rvX, \rvY)
\;,
\label{eq:CE-decomp-mutual}\eeq

\beq
H(\rvX : \rvY) = H(\rvX) -  H(\rvX  |  \rvY)
\;,
\label{eq:CE-info-trans}\eeq

\beq
H(\rvX : \rvY) = H(\rvY) -  H(\rvY  |  \rvX)
\;.
\eeq
In Eq.(\ref{eq:CE-info-trans}), one may think of
$H(\rvX)$ as the information about $\rvX$ prior to 
transmitting it, and $H(\rvX | \rvY)$ as the information about 
$\rvX$ once $\rvX$ is transmitted and $\rvY$ is found out. Since $H(\rvX : \rvY)$
is the difference between the two, one may think of it as
the information (or entropy) ``transmitted" from  $\rvX$ to $\rvY$.
This interpretation of $H(\rvX : \rvY)$ is an alternative to the dependency
interpretation mentioned above.

Let 
$\rvX = (\rvx.)_{\Gamma_1}$,
$\rvY = (\rvx.)_{\Gamma_2}$ and
$\rvZ = (\rvx.)_{\Gamma_3}$, where the
$\Gamma_1, \Gamma_2, \Gamma_3$ are
non-empty, possibly overlapping, subsets of $\zn$.
We can extend further the  domain of the function $H(\cdot)$ by introducing the
following axioms 

\beq
 H[ (\rvX, \rvY): \rvZ] = H[ (\rvX:\rvZ), (\rvY:\rvZ)]
\;,
\label{eq:CE-colon-distri}\eeq

\beq
 H[ (\rvX: \rvY), \rvZ] = H[ (\rvX,\rvZ) : (\rvY,\rvZ)]
\;.
\label{eq:CE-comma-distri}\eeq
Eq.(\ref{eq:CE-colon-distri}) means that ``:" distributes over ``,". According to Eq.(\ref{eq:CE-decomp-mutual}), the LEFT
hand side of Eq.(\ref{eq:CE-colon-distri}) equals $H(\rvX, \rvY) + H(\rvZ) - H(\rvX, \rvY, \rvZ)$.
Eq.(\ref{eq:CE-comma-distri}) means that ``," distributes over ``:". According to Eq.(\ref{eq:CE-decomp-mutual}), the RIGHT
hand side of Eq.(\ref{eq:CE-comma-distri}) equals $H(\rvX, \rvZ) + H(\rvY, \rvZ) - H(\rvX, \rvY, \rvZ)$.
With the help of the above distributive laws, the entropy of a compound expression with any
number of 
``:" and ``$|$" operators can be expressed as a sum of $(\pm 1) H(\cdot)$ functions 
containing ``,"  but not containing ``:" and ``$|$" in their arguments.
For example, if $\rva, \rvb,\rvc$
are nodes of a QB net, then

\beq
\begin{array}{l}
H[ (\rva: \rvb)  |  \rvc] = H[(\rva : \rvb), \rvc] - H(\rvc)\\
\;\;\; = H[(\rva, \rvc): (\rvb, \rvc)] - H(\rvc)\\
\;\;\; = H(\rva, \rvc) + H(\rvb, \rvc) - H(\rva, \rvb, \rvc) - H(\rvc)
\end{array}
\;.
\eeq

If some parentheses are omitted within the argument of $H(\cdot)$,
the argument may become ambiguous. For example, does
$H(\rva : \rvb, \rvc)$ mean $H((\rva : \rvb), \rvc)$ or  $H(\rva : (\rvb, \rvc))$?
Ambiguous arguments should be interpreted using the following operator 
precedence order, from highest to lowest precedence: comma(,), colon(:), vertical line($|$). Thus,
$H(\rva : \rvb, \rvc)$ should be interpreted as $H(\rva : (\rvb, \rvc))$.

In the mathematical field called Set Theory, one defines
the union $A \cup B$, the intersection $A \cap B$ and the 
difference $A-B = A \cap {\rm complement} (B)$ of two sets
$A$ and $B$. One also defines functions $\mu(\cdot)$ called
measures. A {\it measure} $\mu(\cdot)$ assigns a non-negative 
real number to any ``measurable" set $A$. $\mu(\cdot)$ satisfies

\beq
\mu(\emptyset) = 0
\;,
\eeq

\beq
\mu(\cup_{i=0}^{\infty} E_i) = \sum_{i=0}^{\infty} \mu( E_i)
\;,
\eeq
where $\emptyset$ is the empty set, and the $E_i$'s are disjoint measurable sets. For example,
for any set $S = [a_1, b_1] \cup [a_2, b_2] \cup \ldots \cup [a_N, b_N]$,
where the $[a_i,  b_i]$'s are disjoint closed intervals of 
real numbers, one can define $\mu(S) = \sum_{i=0}^N (b_i - a_i)$.

There is a close analogy between the properties of entropy
functions in Information Theory(IT) and those of measure functions in Set Theory(ST).
If $A, B$ are sets and $\rva, \rvb$ are node random variables, then it
is fruitful to imagine the following correspondences\cite{lattice}:

\beq
\begin{array}{llll}
{\rm atoms:} 				& A 		&\longleftrightarrow	& \rva \\
{\rm binary\;operators:} 	& A\cup B 	&\longleftrightarrow	& (\rva, \rvb) \\
							& A\cap B 	&\longleftrightarrow	& (\rva : \rvb) \\
							& A- B 		&\longleftrightarrow	& (\rva  |  \rvb) \\
{\rm real-valued\;function:}  & \mu(A)  &\longleftrightarrow	& H(\rva)
\end{array}
\;.
\eeq
In both ST and IT, one defines a 
real-valued function (i.e., $\mu(\cdot)$ in ST versus $H(\cdot)$ in IT). This
real-valued function takes as arguments certain well-formed
expressions. A well-formed expression consists of either a single atom
(a set in ST versus a node random variable in IT) or a compound expression.
A compound expression is formed by using binary operators
($\;\cup \; \cap \;  -\;$ in ST versus $\;, \;  : \;   | \;$ in IT) 
to bind together either
(1) 2 atoms or (2) an atom and another compound expression or (3) two compound expressions.

Table \ref{Table-ent} gives a list of properties 
(identities and inequalities) satisfied by the classical entropy $H(\cdot)$.
Whenever possible, Table \ref{Table-ent} matches each property of
entropy functions with an analogous property of
measure functions.
See Refs.\cite{Mans}-\cite{SchuLecs} to get  proofs of
those statements in Table \ref{Table-ent} that are not proven in this paper.

\EndSection

\newpage


\addcontentsline{toc}{subsection}{Table 1. Entropy Properties}

\begin{center}
\label{Table-ent}
{\scriptsize
{\bf Table 1. ENTROPY PROPERTIES} (compiled by R.R.Tucci, report errors to tucci@ar-tiste.com)
\begin{tabular}{|l|l|l|}

\hline
	$\mu(A-B) = \mu(A \cup B)  - \mu(B)$  & $H(\rvX  |  \rvY) = H(\rvX, \rvY) - H(\rvY)$& $H\rarrow S_\rho$\\
	($-$ in terms of $\cup$) & & \\
\hline
	$\mu(A\cap B) = \mu(A) + \mu(B) - \mu(A \cup B)$ & $H(\rvX : \rvY) = H(\rvX) + H(\rvY) - H(\rvX, \rvY)$ & $H\rarrow S_\rho$\\
	($\cap$ in terms of $\cup$) & & \\
\hline
	$\mu((A\cup B) \cap C) = \mu( (A\cap C) \cup (B \cap C))$ & 
	$H( (\rvX , \rvY) : \rvZ ) = H( (\rvX : \rvZ), (\rvY : \rvZ) )$& $H\rarrow S_\rho$\\
	($\cap$ distributes over $\cup$) & &\\
\hline
	$\mu((A\cap B) \cup C) = \mu( (A\cup C) \cap (B \cup C))$ & 
	$H( (\rvX : \rvY) , \rvZ ) = H( (\rvX , \rvZ) : (\rvY , \rvZ) )$ & $H\rarrow S_\rho$\\
	($\cup$ distributes over $\cap$) & &\\
\hline \hline	
	$0 \leq \mu(A) $ & $0 \leq H(\rvX) \leq \log_2 N_\rvX$ & $H\rarrow S_\rho$\\
	(non-negative) & $H(\rvX)=0$ iff $P(X_0)=1$ for some $X_0\in S_\rvX$, &Let $\rho' = \tr \rho$, where trace is over\\
	  & and $P(X)=0$ for all other $X\in S_\rvX$. &all random variables except $\rvX$.\\
	  & $H(\rvX)= \log_2 N_\rvX$ iff $P(X) = \frac{1}{N_\rvX}$ for all $X$. & $S_\rho(\rvX)=0$ iff $\rho'$ is pure.\\
	  && $S_\rho(\rvX)= \log_2 N_\rvX$ iff $\rho' = I \frac{1}{N_\rvX}$\\ 
\hline
	$\mu(B) \leq \mu(A\cup B)$ & $H(\rvY) \leq H(\rvX, \rvY)$ & $|S_\rho(\rvX) - S_\rho(\rvY)|\leq S_\rho(\rvX, \rvY)$\\
	or $0 \leq \mu(A-B)$ & or $0 \leq H(\rvX  |  \rvY)$ & Triangle Inequality (Araki-Lieb).\\
	 & Equality iff $\rvX = f(\rvY)$  & $S_\rho(\rvX | \rvY)$ may be negative!\\ 
	 &for some function $f(\cdot)$. & Let $\rho' = \tr \rho$, where trace is over\\
	 && all random variables except $\rvX, \rvY$.\\
	 && Equality iff $\rho'$ is pure. Schmidt \\
	 && Decomp. applies when $\rho'$ is pure.\\
\hline
	$\mu(A\cup B) \leq \mu(A) + \mu(B)$ & $H(\rvX, \rvY) \leq H(\rvX) + H(\rvY)$ & $H\rarrow S_\rho$\\
	or $\mu(A-B) \leq \mu(A)$ & or $H(\rvX  |  \rvY) \leq H(\rvX)$ & Let $\rho' = \tr \rho$, where trace is over\\
	or $0 \leq \mu(A\cap B)$. & or $0 \leq H(\rvX : \rvY)$. &  all random variables except $\rvX, \rvY$.\\
	(sub-additivity) & Equality iff $\rvX$ and $\rvY$ are independent. &Equality iff $\rho' = (\tr_\rvX \rho') (\tr_\rvY \rho')$\\
\hline
	$\mu(A - (B \cup C) ) \leq \mu (A-B) $ & $H(\rvX  |   (\rvY, \rvZ) ) \leq H(\rvX  |  \rvY)$& $H\rarrow S_\rho$\\
	(strong sub-additivity) & & (Lieb-Ruskai)\\
\hline\hline
	&& $S(U \rho U^\dagger) = S(\rho)$,\\
	&& for any unitary matrix $U$.\\
	&&Thus, if $\rho$ has eigenvalues $p_j$,\\
	&&then $S(\rho) = H(\vec{p}).$\\
\hline
	& & $S(\rho) \leq H(\vec{p})$,\\
	& & where $p_i = <i | \rho | i>$. Equality iff\\
	&&  $<i | \rho | j>$ = 0 for all $i\neq j$.\\
\hline
	& & $S\left(\sum_j p_j |j><j|)\right) \leq H(\vec{p})$,\\
	&& where $p_j$ is a prob. distribution.\\
	& & Equality iff $<j  |  j'> = \delta(j, j').$\\
\hline
	& $\sum_j p_j \log_2 \frac{q_j}{p_j} \leq 0$ & $-\tr( \rho (\log_2 \rho - \log_2 \sigma)) \leq 0$\\
	& where $p_j$ and $q_j$ are prob. distributions. & where $\rho, \sigma$ are density matrices.\\
	&Gibbs' inequality. & \\
	& Equality iff $q_j = p_j$ for all $j$. & Equality iff $\rho = \sigma$.\\
\hline
	&$\sum_\alpha w_\alpha H(\vec{p}_\alpha)\leq H(\sum_\alpha w_\alpha \vec{p}_\alpha),$
	&$\sum_\alpha w_\alpha S(\rho_\alpha) \leq S(\sum_\alpha w_\alpha \rho_\alpha),$\\
	&where $w_\alpha\geq 0$ and $\sum_\alpha w_\alpha=1$.
	&where $w_\alpha\geq 0$ and $\sum_\alpha w_\alpha=1$.\\
	&Convexity.
	&Convexity. \\
	&Equality iff $\exists \vec{p}$ such that $\forall \alpha, \vec{p}_\alpha = \vec{p}$.
	&Equality iff $\exists \rho$ such that $\forall \alpha, \rho_\alpha = \rho$.\\
\hline 
	&$ H(\sum_\alpha w_\alpha \vec{p}_\alpha)$
	&$ S(\sum_\alpha w_\alpha \rho_\alpha)$\\
	&$\;\;\;\leq \sum_\alpha w_\alpha H(\vec{p}_\alpha) - \sum_\alpha w_\alpha \log_2 w_\alpha, $
	&$\;\;\;\leq \sum_\alpha w_\alpha S(\rho_\alpha) - \sum_\alpha w_\alpha \log_2 w_\alpha, $\\
	&where $w_\alpha\geq 0$ and $\sum_\alpha w_\alpha=1$. 
	&where $w_\alpha\geq 0$ and $\sum_\alpha w_\alpha=1$. \\
	&Equality iff $\vec{p}_\alpha \cdot \vec{p}_\beta = 0$ for $\alpha \neq \beta$.
	&Equality iff $\rho_\alpha \rho_\beta = 0$ for $\alpha \neq \beta$.\\
	&Equality is Shannon grouping axiom for $H(\cdot)$.& Lanford-Robinson.\\
\hline	
\end{tabular}
}
\end{center}

\EndSection


\BeginSection{CB Net Examples}\label{sec:cbnet-examples}
In Section \ref{sec:c-ent}, we discussed entropic properties
which are valid for all CB nets. In this section, we will discuss entropic properties that apply 
to particular CB nets.

First, we will consider all possible CB nets with 2 and 3 nodes.
Their nodes will be labelled by the random variables $\rva, \rvb, \rvc$.

\begin{figure}[h]
	\begin{center}
	\epsfig{file=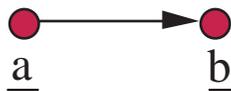}
	\caption{Two connected nodes.}
	\label{fig:2-nds}
	\end{center}
\end{figure}

Fig.(\ref{fig:2-nds}) shows two connected nodes. By the definition of CB nets, the 
joint probability $P(a, b)$ of the two nodes of this net satisfies:

\beq
P(a, b) = P(b  |  a) P(a)
\;.
\eeq
Taking the logarithms and then the expected values of both sides of the last equation
yields

\beq
H(\rva, \rvb) = H(\rvb  |  \rva) + H(\rva)
\;.
\eeq

\begin{figure}[h]
	\begin{center}
	\epsfig{file=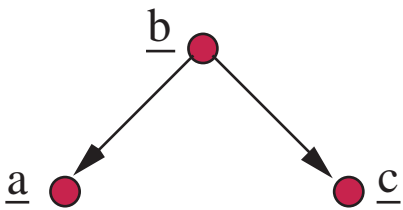}
	\caption{Diverging graph with 3 nodes.}
	\label{fig:div}
	\end{center}
\end{figure}

Fig.(\ref{fig:div}) shows a ``diverging" graph with 3 nodes. By the definition of
CB nets, the joint probability $P(a, b, c)$ of all the nodes of this net satisfies:

\beq
P(a, b, c) = P(b) P(a  |  b) P(c  |  b)
\;.
\eeq
The last equation implies the following entropic constraint:

\beq
\begin{array}{ll}
H(\rva, \rvb, \rvc) &= H(\rvb) + H(\rva  |  \rvb) + H(\rvc  |  \rvb)\\
&=H(\rva, \rvb) + H(\rvc, \rvb) - H(\rvb)
\end{array}
\;,
\eeq
which is equivalent to 

\beq
H[ (\rva : \rvc)  |  \rvb] = 0
\;.
\label{eq:CBE-cond-indep1}\eeq
This means that at a fixed value of $\rvb$, $\rva$ and $\rvc$ are
independent random variables.

\begin{figure}[h]
	\begin{center}
	\epsfig{file=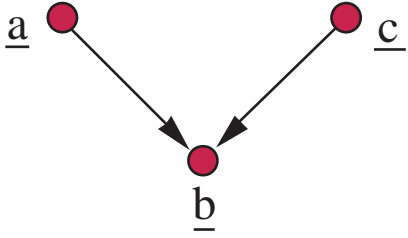}
	\caption{Converging graph with 3 nodes.}
	\label{fig:conv}
	\end{center}
\end{figure}

Fig.(\ref{fig:conv}) shows a ``converging" graph with 3 nodes. $P(a,b,c)$ for this net
must satisfy

\beq
P(a, b, c) = P(a) P(c) P(b  |  a,c)
\;.
\eeq
Thus,

\beq
\begin{array}{ll}
H(\rva, \rvb, \rvc) & = H(\rva) + H(\rvc) + H(\rvb  |  \rva, \rvc)\\
& = H(\rva) + H(\rvc) - H(\rva, \rvc) + H(\rva, \rvb, \rvc)
\;,
\end{array}
\eeq
which is equivalent to 

\beq
H(\rva : \rvc) = 0
\;.
\eeq
This means that $\rva$ and $\rvc$ are independent.

\begin{figure}[h]
	\begin{center}
	\epsfig{file=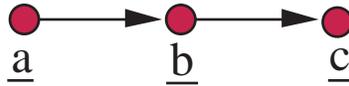}
	\caption{Three node Markov chain.}
	\label{fig:markov}
	\end{center}
\end{figure}

A Bayesian net consisting of a simple chain of $N$ nodes connected 
by arrows all pointing in the same direction 
will be called an {\it $N$ node Markov chain}. If the nodes are labelled by random
variables $\rvq_1, \rvq_2, \cdots, \rvq_N$, we will denote the net by 
$\rvq_1 \rarrow \rvq_2 \rarrow \cdots \rarrow \rvq_N$.
Fig.(\ref{fig:markov}) shows a 3 node Markov chain
$\rva \rarrow \rvb \rarrow \rvc$. $P(a, b, c)$ for this net
must satisfy:

\beq
P(a, b, c) = P(c  |  b) P(b  |  a) P(a)
\;.
\eeq
Thus,

\beq
\begin{array}{ll}
H(\rva, \rvb, \rvc) &= H(\rvc  |  \rvb) + H(\rvb  |  \rva) + H (\rva)\\
&=H(\rvc, \rvb) - H(\rvb) + H(\rvb, \rva)
\;,
\end{array}
\eeq
which is equivalent to 

\beq
H[ (\rva : \rvc)  |  \rvb] = 0
\;.
\label{eq:CBE-cond-indep2}\eeq
Note that Eq.(\ref{eq:CBE-cond-indep2}) for the Markov chain Fig.(\ref{fig:markov}) 
is the same as Eq.(\ref{eq:CBE-cond-indep1})
for the diverging graph Fig.(\ref{fig:div}). This shows that two CB nets with different topologies
can have the same entropic constraint.

\begin{figure}[h]
	\begin{center}
	\epsfig{file=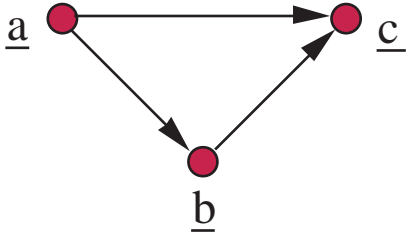}
	\caption{Fully connected 3 node graph.}
	\label{fig:3-full}
	\end{center}
\end{figure}

Fig.(\ref{fig:3-full}) shows a fully connected 3 node graph. $P(a, b, c)$ for this net
must satisfy:

\beq
P(a, b, c) = P(c  |  b, a) P(b  |  a) P(a)
\;.
\label{eq:CBE-p-taut}\eeq
Because the graph is fully connected, Eq.(\ref{eq:CBE-p-taut}) is a tautology: it
is satisfied by all probability distributions $P(a, b, c)$. Eq.(\ref{eq:CBE-p-taut})
implies

\beq
H(\rva, \rvb, \rvc) = H(\rvc  |  \rvb, \rva) + H(\rvb  |  \rva) + H(\rva)
\;.
\label{eq:CBE-h-taut}\eeq

\begin{figure}[h]
	\begin{center}
	\epsfig{file=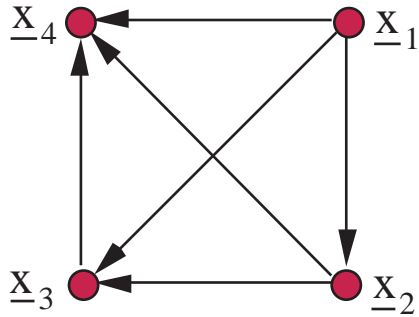}
	\caption{Fully connected 4 node graph.}
	\label{fig:4-full}
	\end{center}
\end{figure}

Eq.(\ref{eq:CBE-h-taut}) can be easily generalized to any number $N\geq 2$ of nodes. Consider a
fully connected CB net with $N$ nodes labelled by
the random variables $\rvx_1, \rvx_2, \ldots , \rvx_N$.
Fig.(\ref{fig:4-full}) shows the case $N=4$. By the definition of CB nets, the joint probability 
of all the nodes must satisfy:

\beq
P(x_1, x_2, \ldots, x_N) =
\prod_{j=1}^N P(x_j  |  x_{j-1}, \ldots , x_2, x_1)
\;.
\eeq
Thus,

\beq
H(\rvx_1, \rvx_2, \ldots, \rvx_N) =
\sum_{j=1}^N H(\rvx_j  |  \rvx_{j-1}, \ldots , \rvx_2, \rvx_1)
\;.
\eeq

Consider a 3 node Markov chain 
 $\rvq_1 \rarrow \rvq_2 \rarrow \rvq_3$.
We shall demonstrate that:

\beq
0 = H(\rvq_1  |  \rvq_1) \leq
H(\rvq_1  |  \rvq_2) \leq
H(\rvq_1  |  \rvq_3)
\;,
\label{eq:CBE-dp-ineq-cond}\eeq
and 

\beq
H(\rvq_1) = H(\rvq_1 : \rvq_1)\geq
H(\rvq_1 : \rvq_2) \geq
H(\rvq_1 : \rvq_3)
\;.
\label{eq:CBE-dp-ineq-mut}\eeq
Eqs.(\ref{eq:CBE-dp-ineq-cond}) and (\ref{eq:CBE-dp-ineq-mut}) will be called
{\it fixed sender (or speaker) data processing (DP) inequalities}.
Eq.(\ref{eq:CBE-dp-ineq-cond}) tells us that
the entropy of $\rvq_1$ increases as ``time" increases, because 
the ``memory" $\rvq_j$ of $\rvq_1$ becomes a progressively less faithful representation of the original.
Eq.(\ref{eq:CBE-dp-ineq-mut}) tells us that
the dependency of $\rvq_j$ on $\rvq_1$
decreases as ``time" $j$ increases. 
Alternatively, one might say that the amount of information 
transmitted from $\rvq_1$ to $\rvq_j$ decreases as the ``distance" $j$ increases, 
The farther away the receiver is from the sender, the less  
information it gets.
Eq.(\ref{eq:CBE-dp-ineq-mut}) follows trivially from Eq.(\ref{eq:CBE-dp-ineq-cond})
Just subtract $H(\rvq_1)$
from each term of Eq.(\ref{eq:CBE-dp-ineq-cond}) and multiply the whole string of inequalities by $-1$.
To prove Eq.(\ref{eq:CBE-dp-ineq-cond}),
we begin by noticing that 

\beq
P(q_1  |  q_2, q_3) =
\frac{P(q_3 |  q_2 )P(q_2  |  q_1) P( q_1)}{ \sum_{q_1'} P(q_3 |  q_2 )P(q_2  |  q_1') P( q_1')}=
\frac{P(q_2  |  q_1) P( q_1)}{ \sum_{q_1'} P(q_2  |  q_1') P( q_1')}=
P(q_1  |  q_2)
\;.
\label{eq:CBE-p-redux}\eeq
This just means that once $\rvq_2$ is known, finding out $\rvq_3$ adds nothing new to our
knowledge of $\rvq_1$. Eq.(\ref{eq:CBE-p-redux}) implies 

\beq
H(\rvq_1  |  \rvq_2, \rvq_3) = H(\rvq_1 | \rvq_2)
\;.
\eeq
Using the last equation and strong sub-additivity, we obtain

\beq
H(\rvq_1  |  \rvq_2)  =
H(\rvq_1  |  \rvq_2, \rvq_3) \leq  H(\rvq_1  |  \rvq_3)
\;.
\eeq
QED.

The Markov chain 
 $\rvq_1 \rarrow \rvq_2 \rarrow \rvq_3$
also satisfies

\beq
H(\rvq_3  |  \rvq_2) \leq
H(\rvq_3  |  \rvq_1)
\;,
\label{eq:CBE-dp-ineq-cond-f-rec}\eeq
and 

\beq
H(\rvq_3 : \rvq_2) \geq
H(\rvq_3 : \rvq_1)
\;.
\label{eq:CBE-dp-ineq-mut-f-rec}\eeq
Eqs.(\ref{eq:CBE-dp-ineq-cond-f-rec}) and (\ref{eq:CBE-dp-ineq-mut-f-rec}) will be called
{\it fixed receiver (or listener) data processing (DP) inequalities}.
As in the fixed sender case,
Eq.(\ref{eq:CBE-dp-ineq-mut-f-rec}) follows trivially from Eq.(\ref{eq:CBE-dp-ineq-cond-f-rec}).
Just subtract $H(\rvq_3)$
from each term of the inequality and multiply by $-1$.
To prove Eq.(\ref{eq:CBE-dp-ineq-cond-f-rec}),
we first realize that
the method employed in Eq.(\ref{eq:CBE-p-redux}) can be used to show that

\beq
P(q_3  |  q_2, q_1) =
P(q_3  |  q_2)
\;.
\label{eq:CBE-p-redux-f-rec}\eeq
Whereas in the fixed sender case, Eq.(\ref{eq:CBE-p-redux}) 
told us that we need only condition on the closest of the later times, 
Eq.(\ref{eq:CBE-p-redux-f-rec}) instructs us to condition only on the closest of the earlier times.
Eq.(\ref{eq:CBE-p-redux-f-rec}) implies 

\beq
H(\rvq_3  |  \rvq_2, \rvq_1) = H(\rvq_3 | \rvq_2)
\;.
\eeq
Using the last equation and strong sub-additivity, we obtain

\beq
H(\rvq_3  |  \rvq_2)  =
H(\rvq_3  |  \rvq_2, \rvq_1) \leq  H(\rvq_3  |  \rvq_1)
\;.
\eeq
QED.

Eqs.(\ref{eq:CBE-dp-ineq-mut}) and (\ref{eq:CBE-dp-ineq-mut-f-rec}) 
can be stated simultaneously as

\beq
H(\rvq_1 : \rvq_3) \leq \min \{ H(\rvq_1 : \rvq_2), H(\rvq_2, \rvq_3) \}
\;.
\eeq

Consider the  4 node Markov chain 
$\rvq_1 \rarrow \rvq_2 \rarrow \rvq_3 \rarrow \rvq_4$. Then

\beq
H(\rvq_1 : \rvq_4) \leq H(\rvq_2 : \rvq_3)
\;.
\eeq 
This follows from

\beq
H(\rvq_1 : \rvq_4) \leq H(\rvq_1 : \rvq_3) \leq H(\rvq_2 : \rvq_3)
\;,
\eeq
where we have used the fixed sender DP inequality first and the 
fixed receiver DP inequality second.

It is also interesting to note that the fixed receiver and fixed sender DP inequalities
are related by time reversal. Indeed,
suppose we are given a 3 node Markov chain 
$\rvq_1 \rarrow \rvq_2 \rarrow \rvq_3$. Then we can extend it to a 
5 node Markov chain  $\rvq_1 \rarrow \rvq_2 \rarrow \rvq_3 \rarrow \rvq_2' \rarrow \rvq_1' $.
We need to define the set of states and the transition matrices for nodes $\rvq_2'$ and $\rvq_1'$.
Suppose we do this as follows:

\begin{subequations}
\beq
S_{\rvq_2'} = S_{\rvq_2}
\;,
\eeq

\beq
S_{\rvq_1'} = S_{\rvq_1}
\;,
\eeq
\end{subequations}

\begin{subequations}
\beq
P(\rvq_2' = q_2 | \rvq_3 = q_3) = P(\rvq_2 = q_2 | \rvq_3 = q_3) =
\frac{ \sum_{q_1} P(q_1, q_2, q_3) }{ \sum_{q_1, q_2} P(q_1, q_2, q_3) }
\;,
\eeq

\beq
P(\rvq_1' = q_1 | \rvq_2' = q_2) = P(\rvq_1 = q_1 | \rvq_2 = q_2) =
\frac{ \sum_{q_3} P(q_1, q_2, q_3) }{ \sum_{q_1, q_3} P(q_1, q_2, q_3) }
\;,
\eeq
\end{subequations}
where

\beq
P(q_1, q_2, q_3) = P(q_3| q_2) P(q_2 | q_1) P(q_1)
\;.
\eeq
Then, applying the fixed sender DP inequality leads to the fixed receiver one: 

\beq
H(\rvq_3 : \rvq_2) = H(\rvq_3 : \rvq_2') \geq H(\rvq_3 : \rvq_1') = H(\rvq_3 : \rvq_1)
\;.
\eeq

Can the DP inequalities,
which are reminiscent of the Second Law of Thermodynamics, be generalized 
easily and naturally to Bayesian nets more 
complicated than merely Markov chains? Such a generalization could  
turn out to be very  useful. After all,
the Second Law of Thermodynamics is 
an extremely useful result. See  
\cite{DP_Ineq} for a generalization.

\EndSection


\BeginSection{Reduced Density Matrices}\label{sec:red-dmats}
In preparation for the next section, we will show in this section how to use
a density matrix to generate a new, ``reduced" density matrix. The
Hilbert space acted upon by the reduced density matrix will have 
smaller dimension than the Hilbert space acted upon by the 
progenitor density matrix.

Recall that a density matrix is an operator $\rho$ 
acting on a Hilbert space $\cal H$. In addition, $\rho$ must be a
Hermitian operator with unit trace and non-negative eigenvalues.
An operator with non-negative eigenvalues is called a {\it non-negative (or positive indefinite) operator}.
Note that if $\sigma$ is a Hermitian operator that acts on
a Hilbert space $\cal H$, then $\sigma$ has non-negative
eigenvalues iff $\bra{\phi} \sigma \ket{\phi}\geq 0$ for all $\ket{\phi}\in \cal H$.
This is why. Let's represent $\sigma$ by a matrix and the elements of 
$\cal H$ by column vectors. Matrix $\sigma$ can be expressed as
$\sigma = U \Lambda U^\dagger$, where $U$ is a unitary matrix and
$\Lambda$ is a diagonal matrix whose diagonal entries $\lambda_i$ are the eigenvalues of $\sigma$.
 If $\phi$ is any vector in $\cal H$, and 
$v_i$ are the components of vector $v = U^\dagger \phi$, then 

\beq
\phi^\dagger \sigma \phi = \sum_i |v_i|^2 \lambda_i
\;.
\eeq
From the last equation, it is clear that $\phi^\dagger \sigma \phi \geq 0$ 
for all $\ket{\phi}\in \cal H$ iff $\lambda_i \geq 0$ for all $i$. 

For any operator $\sigma$
acting on $\hila$ and for which $\trace{\rva}\sigma \neq 0$, 
it is convenient to define the normalizing function ${\cal N}(\sigma)$  by

\beq
{\cal N}(\sigma) = \frac{\sigma}{\trace{\rva} \sigma }
\;.
\eeq

Now suppose that $\rho$ is a density matrix acting on $\hila \otimes \hilb$ , and
$\pi_\rva$ is a projection operator ($\pi_\rva^2 = \pi_\rva$) acting on $\hila$.
Let 

\beq
K  = \trace{\rva, \rvb} (\pi_\rva \rho)
\;.
\eeq
If we define

\beq
\ket{\phi_{ab}} = (\pi_\rva \ket{a})\ket{b}  
\;
\eeq
for all $a\in S_\rva$ and $b\in S_\rvb$, then

\beq
K = \sum_{a,b} \bra{\phi_{ab}} \rho \ket{\phi_{ab}} \geq 0
\;.
\eeq
When $K\neq 0$, we can define the {\it reduced density matrix} 
$\redmat{\pi_\rva}(\rho)$ by

\beq
\redmat{\pi_\rva} (\rho)= {\cal N} [\trace{\rva}( \pi_\rva \rho) ]=
K^{-1}\trace{\rva} (\pi_\rva \rho)
\;.
\eeq
Note that $\redmat{\pi_\rva}(\rho)$ is indeed a density matrix.
Clearly, it is Hermitian and it has unit trace. Furthermore,
for any $\ket{\beta}\in \hilb$, if we define 

\beq
\ket{\chi_{a\beta}} = (\pi_\rva \ket{a})\ket{\beta}
\;
\eeq
for all $a \in S_\rva$, then 

\beq
\bra{\beta} \redmat{\pi_\rva} (\rho) \ket{\beta} =
K^{-1} \sum_a \bra{\chi_{a\beta}} \rho \ket{\chi_{a\beta}} \geq 0
\;.
\eeq

Some possibilities for $\pi_\rva$ are: 
\begin{description}
\item[(a)] $\pi_\rva=1$. Then
	\beq
	\redmat{\pi_\rva} \rho = \trace{\rva} \rho
	\;.
	\eeq
	Note that $\trace{\rva} (U \rho U^\dagger) = \trace{\rva}(\rho)$
	for any unitary matrix $U$ acting on $\hila$, However, for other $\pi_\rva$'s,
	it may happen that
	$\redmat{\pi_\rva} (U \rho U^\dagger) \neq \redmat{\pi_\rva}(\rho)$. Thus,
	 although not true for $\trace{\rva}(\cdot)$,
	 $\redmat{\pi_\rva}(\cdot)$  may depend
	 on the basis used to evaluate it.
\item[(b)] $\pi_\rva = \ket{\alpha}\bra{\alpha}$, where $\ket{\alpha}\in \hila$. Then
	\beq
	\redmat{\pi_\rva} \rho = 
	\frac{
	\bra{\alpha} \rho \ket{\alpha}
	}{
	\bra{\alpha} \trace{\rvb}(\rho) \ket{\alpha}
	}
	\;.
	\eeq
If $a, a'\in S_\rva$, then some possibilities for $\ket{\alpha}$ are
$\ket{a}$, $\frac{1}{\sqrt{2}}(\ket{a} + \ket{a'})$, and  $\ket{Av(\rva)}$,
where 

\beq
\ket{Av(\rva)} = \frac{1}{\sqrt{N_\rva}} \sum_{a\in S_\rva}\ket{a}
\;.
\eeq
We will call $\ket{Av(\rva)}$ the {\it average of the  $\rva$  basis}.
\end{description}
Define

\beq
E_\rva = \ket{Av(\rva)} \bra{Av(\rva)}
\;,
\eeq

\beq
K = \bra{Av(\rva)} \trace{\rvb} (\rho) \ket{Av(\rva)}
\;.
\eeq
If $K\neq 0$, we can 
define  the {\it entry sum  $\esum{\rva}(\rho)$ of  $\rho$ in the $\rva$ basis } 
by

\beq
\esum{\rva}(\rho) = \redmat{E_\rva} (\rho)
\;.
\eeq
Thus, 
\beq 
\esum{\rva}(\rho) = {\cal N} [\trace{\rva}(E_\rva\rho)] = K^{-1} \bra{Av(\rva)} \rho \ket{Av(\rva)}
\;.
\eeq
$\esum{\rva}(\rho)$ is called an entry sum because it can be
expressed as

\beq
\esum{\rva}(\rho) = {\cal N} (\sum_{a_1, a_2} \bra {a_1} \rho \ket{a_2})
\;,
\eeq
where the sum is over all $a_1\in S_\rva$ and $a_2\in S_\rva$.

\EndSection


\BeginSection{Density Matrices Associated with a QB Net}\label{sec:assoc-dmats}

In this section, we will describe a method for constructing 
many different density matrices
associated with a single
QB net.

Consider a QB net  $\qnet$ with $N$ nodes
labelled by the random variables $\rvx_1, \rvx_2, \ldots, \rvx_N$.

We will consider density matrices which act on 
$\hil_{(\rvx.)_\Gamma}$, where $\Gamma$ is a subset of $\zn$. We will
use $\Gamma(\rho)$ to represent the $\Gamma$ of density matrix $\rho$.

Let $A(x_.)$ be the amplitude assigned by  $\qnet$ to story $x_\cdot$ .
Assume that (see Appendix \ref{app:bnet-review})

\beq
\sum_{x.} \left | A(x.) \right|^2 = 1
\;.
\label{eq:ADM-meta-norm}\eeq
Then we can define the {\it meta state-vector} $\ketmeta$ and
the {\it meta density matrix} $\mu$ of  $\qnet$ by

\beq
\ketmeta = \sum_{x.} A(x.) \ket{x.}
\;,
\label{eq:ADM-ketmeta}\eeq

\beq
\mu = \ketmeta \bra{\psi_{meta}}
\;.
\eeq
(Eq.(\ref{eq:ADM-meta-norm}) guarantees that $\ketmeta$ has unit magnitude.)
For example, if  $\qnet$ has 3 nodes $\rva, \rvb, \rvc$, then

\beq
\ketmeta = \sum_{a,b.c} A(a, b, c) \ket{a,b,c}
\;,
\eeq

\beq
\mu = \sum_{ri} A(a, b, c) A^*(a', b', c')
\ket{a, b, c}
\bra{a', b', c'}
\;.
\label{eq:ADM-eg-mu}\eeq

Note that $\ket{x.}$ in Eq.(\ref{eq:ADM-ketmeta}) represents a ket in the Hilbert space 
$\hil_{\rvx.} = \hil_{\rvx_1} \otimes \hil_{\rvx_2} \otimes \ldots\otimes \hil_{\rvx_N}$.
This is not the conventional use of a tensor product of Hilbert spaces.
In Quantum Mechanics, such products are conventionally used to represent a ``system" described
by $\hil_{\rvx.}$ which consists of $N$ ``subsystems" such that the i'th subsystem 
is described by $\hil_{\rvx_i}$. 
($\rvx_1$ might correspond to the position and $\rvx_2$ to the spin of
the same particle, so the two subsystems may be associated with the same particle.)
In our usage, the spaces $\hil_{\rvx_i}$ 
correspond to the nodes
of a QB net.
They need not correspond to separate subsystems. 
They might, for example, correspond to the same subsystem at two different times.

Because it acts on this unusual Hilbert space,
the meta density matrix $\mu$  is
unconventional.  So why use it? Because it is uncontestably a density
matrix in the formal sense (Hermitian, unit trace, non-negative.)
Furthermore, as we shall see in what follows, $\mu$
proves to be a very useful tool for discussing QB nets. The reason why 
$\mu$ is so useful is not hard to see.
$\mu$ is a vast storehouse of information about its QB net $\qnet$. In fact,
it stores the amplitude of all the Feynman stories of  $\qnet$.
Applying to $\mu$ one or more $\redmat{} ()$ operators 
of the type discussed in Section \ref{sec:red-dmats}, 
we can generate many different reduced density matrices,
all pertaining to the same QB net  $\qnet$.
For example, for a QB net with 10 nodes, we might consider
$\esum{\rvx_4} \trace{\rvx_2, \rvx_3}\bra{\rvx_1}\mu\ket{\rvx_1}$.

Suppose $\rva$ is one of the nodes $\rvx_j$ of the QB net, and consider
$\redmat{\pi_\rva}\mu$ for various $\pi_\rva$.

\begin{description}
\item[(a)] $\pi_\rva = \ket{a}\bra{a}$ for some $a\in S_\rva$. Then
$\redmat{\pi_\rva}\mu = {\cal N}(\bra{a} \mu \ket{a})$. This 
corresponds to an experiment in which node $\rva$ is measured,
and found to have a particular value $a$. The experiment is run
repeatedly, and those runs for which $\rva\neq a$ are rejected.

\item[(b)] $\pi_\rva = 1$. Then
$\redmat{\pi_\rva}\mu = \trace{\rva} \mu$.  This 
corresponds to an experiment in which node $\rva$ is measured
without any expectations as to the value obtained.  The experiment is run
repeatedly. We sum over the various outcomes of
the $\rva$ measurement. 

\item[(c)] $\pi_\rva = \ket{Av(\rva)}\bra{Av(\rva)}$.
Then $\redmat{\pi_\rva}\mu = \esum{\rva} \mu$. This corresponds
to an experiment in which node $\rva$ is NOT measured.
\end{description}

Suppose $\rho$ is a density matrix 
obtained by reducing a meta density matrix
$\mu$, and suppose $\rho$ acts on  $\hil_{(\rvx.)_{\Gamma(\rho)}}$.
Any node $\rva$ in $(\rvx.)_{\Gamma(\rho)}$ will be said to be {\it uncommitted},
neither measured nor unmeasured.
Any node $\rva$ in $(\rvx.)_{\zn -\Gamma(\rho)}$
will be said to be either measured or unmeasured. It is {\it unmeasured} iff
to go from $\mu$ to $\rho$, 
one of the reductions we performed  was $\redmat{\pi_\rva} = \esum{\rva}$ 
as in case (c) above. If node $\rva$ is measured 
as in case (b) above (i.e., $\redmat{\pi_\rva} = \tr_\rva$), we will say that it has been 
{\it measured passively}. We describe this  measurement as passive because
it does not involve
data rejection by the observer like case (a) above.

Note that external nodes are always measured.
If an observer does not measure them, they are still measured passively by the environment.
Thus, if $\rva$ is an external node, then $\esum{\rva}(\mu)$ cannot be 
realized physically because $\esum{\rva}(\mu)$ describes a situation in which 
$\rva$ is not measured.

Suppose $\rho_{out}$ is obtained by e-summing $\mu$ over all
internal nodes of the graph:

\beq
\rho_{out} = \esum{(\rvx.)_\zin} (\mu)
\;.
\eeq
Then $\rho_{out}$ is a pure state. Here is why.
Define 

\beq
\ket{\psi} = \sum_{x.} A(x.) \ket{(x.)_\zex}
\;.
\eeq
Now note that

\beq
\av{\psi |  \psi} = \sum_{(x.)_\zex}  \left| \sum_{(x.)_\zin} A(x.) \right | ^2 = 1
\;,
\eeq
and

\beq
\ket{\psi}\bra{\psi} =
\sum_{x.}\sum_{x.'}
A(x.) A^*(x.') \ket{(x.)_\zex}\bra{(x.')_\zex} = \rho_{out}
\;.
\eeq
QED. 

\begin{figure}[h]
	\begin{center}
	\epsfig{file=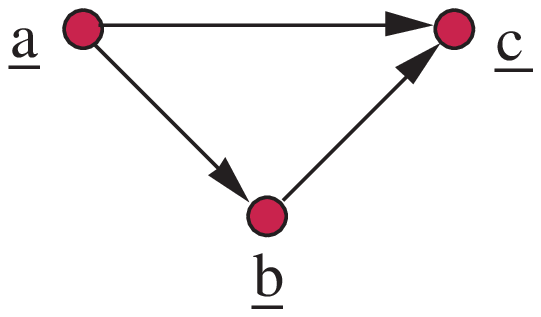}
	\caption{Fully connected 3 node graph.}
	\label{fig:assoc-dmatsF1}
	\end{center}
\end{figure}

To illustrate the definition of $\rho_{out}$, consider Fig.(\ref{fig:assoc-dmatsF1}),
which shows  a fully connected
3 node graph with 
nodes $\rva, \rvb, \rvc$. Nodes $\rva, \rvb$ are internal and 
$\rvc$ is external. 
The $\mu$ for this net is given by Eq.(\ref{eq:ADM-eg-mu}). Define $\rho_{out}$ by

\beq
\rho_{out} = \esum{\rva, \rvb}\mu
\;.
\eeq
If
\beq
\ket{\psi} = \sum_{a,b, c} A(a,b,c) \ket{c}
\;,
\eeq
then
\beq
\ket{\psi}\bra{\psi} = \sum_{all} A(a,b,c) A^*(a',b',c') \ket{c}\bra{c'} = \rho_{out}
\;.
\eeq

$\rho_{out}$ corresponds to a situation in which 
none of the internal nodes are measured and all the external ones are uncommitted. 
We will say that a density matrix has {\it maximum internal coherence} if it corresponds to a situation in which 
none of the internal nodes are measured. $\rho_{out}$ has maximum internal coherence.  
Reduced density matrices obtained by reducing $\rho_{out}$ also have maximum internal coherence.

\EndSection


\BeginSection{Probabilities Associated with a QB Net}\label{sec:assoc-probs}
In this section, we will define various  
probability distributions associated  with a
QB net.

Consider a QB net  $\qnet$ with $N$ nodes
labelled by the random variables $\rvx_1, \rvx_2, \ldots, \rvx_N$.
Let $A(x_.)$ be the amplitude assigned by  $\qnet$ to story $x_\cdot$ .
Suppose $\Gamma$ is a non-empty subset of $ \zn$. The probability of observing
$(\rvx.)_{\Gamma}$ to have a value of $(x.)_{\Gamma}$
 is

\beq
P[ (x.)_{\Gamma}] = 
\frac{ \chi[ (x.)_{\Gamma}]}
{\sum_{(y.)_{\Gamma}} \chi[ (y.)_{\Gamma} ]}
\;,
\label{eq:PA-def}\eeq
where

\beq
\chi[ (x.)_{\Gamma} ] =
\sum_{(x.)_{\zex-\Gamma}} 
\left |
\sum_{(x.)_{\zin-\Gamma}} 
A(x.)
\right |^2
\;.
\label{eq:PA-def-xi}\eeq
In Eq.(\ref{eq:PA-def-xi}) we sum the amplitudes over all internal nodes except
those in $\Gamma$, then we take the magnitude squared, then
we sum that over all external nodes except
those in $\Gamma$.
We can express $P[ (x.)_{\Gamma}]$ in terms of the
meta density matrix of the QB net:

\beq
P[ (x.)_{\Gamma}] = 
\bra{ (x.)_{\Gamma} }
\left(
\trace{(\rvx.)_{\zex-\Gamma}}
\esum{(\rvx.)_{\zin-\Gamma}}
\mu
\right)
\ket{ (x.)_{\Gamma} }
\;.
\label{eq:PA-meta-def}\eeq
Thus, $P[ (x.)_{\Gamma}]$ corresponds to a situation in which the nodes in $\Gamma$ are 
projected to a single state, those in $\zex - \Gamma$ are passively measured, and
those in $\zin - \Gamma$ are not measured at all.
Note that

\beq
\sum_{(x.)_{\Gamma}}  P[ (x.)_{\Gamma}] = 1
\;,
\eeq
as required for a probability distribution. However,
if $\Gamma$ and $\Gamma'$ are non-empty disjoint subsets of $\zn$, then 
it is possible that 

\beq
\sum_{(x.)_{\Gamma'}} 
P[ (x.)_{\Gamma \cup \Gamma'}]
\neq
P[ (x.)_{\Gamma}]
\;.
\eeq

To illustrate the above definition of $P[ (x.)_{\Gamma}]$,
consider the 3 node Markov chain $\rva \rarrow \rvb \rarrow \rvc$. Assume node $\rva$
has amplitudes $\psi_a$, where $\sum_a | \psi_a |^2 = 1$.
Node $\rvb$ (ditto, $\rvc$) has amplitudes $U_{ba}$ (ditto, $V_{cb}$),
where $U_{ba}$ (ditto, $V_{cb}$) are the entries of a unitary matrix. Then

\beq
P(a) = \sum_c \left| \sum_b V_{cb} U_{ba} \psi_a \right |^2  = |\psi_a|^2
\;,
\eeq

\beq
P(b) = \sum_c \left| \sum_a V_{cb} U_{ba} \psi_a \right |^2  = \left|\sum_a U_{ba} \psi_a \right |^2
\;,
\eeq

\beq
P(c) = \left| \sum_{a,b} V_{cb} U_{ba} \psi_a \right |^2 
\;,
\eeq

\beq
P(b,c) = \frac{
\left| \sum_a V_{cb} U_{ba} \psi_a \right |^2
}{
\sum_{b,c}\left| \sum_a V_{cb} U_{ba} \psi_a \right |^2
}
=
\left| \sum_a V_{cb} U_{ba} \psi_a \right |^2
\;,
\eeq

\beq
P(a,b,c) = 
\left| V_{cb} U_{ba} \psi_a \right |^2
\;.
\eeq
Note that 

\beq
\sum_{b,c} P(b,c) = 1
\;,
\eeq
but

\beq
\sum_b P(b,c) \neq P(c)
\;.
\eeq

We can define conditional probabilities using the unconditional ones $P[ (x.)_{\Gamma}]$ defined above.
Suppose $\Gamma_1$ and $\Gamma_2$ are non-empty disjoint subsets of $\zn$. The conditional 
probability 
$P[ (x.)_{\Gamma_1}  |  (x.)_{\Gamma_2}]$
of 
observing
$(\rvx.)_{\Gamma_1}$ to have a value of $(x.)_{\Gamma_1}$,
given or conditioned upon 
the fact that 
$(\rvx.)_{\Gamma_2}$ is known to have the value $(x.)_{\Gamma_1}$,
is

\beq
P[ (x.)_{\Gamma_1}  |  (x.)_{\Gamma_2}]
=
\frac{
P[ (x.)_{\Gamma_1 \cup \Gamma_2}]
}{
P_{\av{(\rvx.)_{\Gamma_1}}} [(x.)_{\Gamma_2}]
}
\;,
\label{eq:PA-def-cond-p}\eeq
where
the denominator of this expression 
is defined by

\beq
P_{\av{(\rvx.)_{\Gamma_1}}} [(x.)_{\Gamma_2}]
=
\sum_{(y.)_{\Gamma_1}}
P[ (y.)_{\Gamma_1} , (x.)_{\Gamma_2}]
\;.
\eeq
Note that

\beq
\sum_{(x.)_{\Gamma_1}}
P[ (x.)_{\Gamma_1}  |  (x.)_{\Gamma_2}]
=1
\;.
\eeq
However,
if $\Gamma_1$, $\Gamma'_1$ and $\Gamma_2$
are non-empty disjoint subsets of $\zn$, then 
it is possible that 

\beq
\sum_{(x.)_{\Gamma'_1}}
P[ (x.)_{\Gamma_1 \cup \Gamma'_1}  |  (x.)_{\Gamma_2}]
\neq
P[ (x.)_{\Gamma_1}  |  (x.)_{\Gamma_2}]
\;.
\eeq

To illustrate the definition of $P[ (x.)_{\Gamma_1}  |  (x.)_{\Gamma_2}]$,
consider again the 3 node Markov chain $\rva \rarrow \rvb \rarrow \rvc$. One has

\beq
P(a,b  |  c) = 
\frac{ 
P(a, b, c)
}{
P_{\av{\rva, \rvb}}(c)
}
\;,
\eeq
where

\beq
P_{\av{\rva, \rvb}}(c) =
\sum_{a, b} P(a,b,c)
\;.
\eeq
Note that

\beq
\sum_{a,b} P(a,b  |  c) =1
\;,
\eeq
but 

\beq
\sum_a P(a,b  |  c) \neq P(b  |  c)
\;.
\eeq

We can easily extend the definition Eq.(\ref{eq:PA-def-cond-p})
of $P[ (x.)_{\Gamma_1}  |  (x.)_{\Gamma_2}]$
to the case that $\Gamma_1$ and $\Gamma_2$ overlap. 
We simply equate $P[ (x.)_{\Gamma_1}  |  (x.)_{\Gamma_2}]$ to 
$P[ (x.)_{\Gamma_1-\Gamma_2}  |  (x.)_{\Gamma_2}]$,
and evaluate the latter with definition Eq.(\ref{eq:PA-def-cond-p}). For example,
for a QB net with nodes $\rva, \rvb, \rvc$,
$P[(a,b)  |  (b,c)] = P[a |  (b, c)]$, and the right-hand side 
can be evaluated with  Eq.(\ref{eq:PA-def-cond-p}).

Given any density matrix associated with the QB net  $\qnet$,
it is natural to define a probability distribution
with its diagonal entries.
Suppose $\rho$ is a density matrix 
that acts on the Hilbert space 
$\hil_{(\rvx.)_\Gamma(\rho)}$, and 
suppose $\Gamma $ is a non-empty subset of $\Gamma(\rho)$.
We define

\beq
P_\rho[ (x.)_\Gamma] =
\bra{ (x.)_\Gamma} \trace{(\rvx.)_{\Gamma(\rho) - \Gamma}}(\rho) \ket{ (x.)_\Gamma}
\;.
\eeq
In the last equation, we trace $\rho$ over all nodes except those contained in $\Gamma$,
then we take the diagonal entries of the resulting operator. 
Note that

\beq
\sum_{(x.)_{\Gamma}}  P_\rho[ (x.)_{\Gamma}] = 1
\;.
\eeq
Furthermore,
if $\Gamma$ and $\Gamma'$ are non-empty disjoint subsets of $\Gamma(\rho)$, then

\beq
\sum_{(x.)_{\Gamma'}} 
P_\rho[ (x.)_{\Gamma \cup \Gamma'}]
=
P_\rho[ (x.)_{\Gamma}]
\;.
\eeq
We can describe the last result by saying that 
the family of probability distributions 
$\{ P_\rho [ (x.)_{\Gamma}]  |  \Gamma \subset \Gamma(\rho)\}$
is closed under marginalization. We saw previously that
the family $\{ P [(x.)_{\Gamma}]  |  \Gamma \subset \zn\}$
does not possess this closure property.

To illustrate the definition of $P_\rho[ (x.)_\Gamma]$,
consider a density matrix $\rho$ which acts on $\hil_{\rva, \rvb, \rvc}$.
Then

\beq
P_\rho(b,c) = \bra{b,c} \trace{\rva} (\rho) \ket{b,c}
\;,
\eeq

\beq
\sum_{b,c} P_\rho(b,c) =1
\;,
\eeq

\beq
\sum_b P_\rho(b, c) = \bra{c} \trace{\rva, \rvb} (\rho) \ket{c} = P_\rho(c)
\;.
\eeq

Note that for any probability distribution $P[(x.)_{\Gamma}]$, we can find a 
density matrix $\rho$ such that

\beq
P [(x.)_{\Gamma}] = P_\rho [(x.)_{\Gamma}]
\;.
\eeq
 Indeed,
just set 

\beq
\rho = \trace{(\rvx.)_{\zex - \Gamma}} [\esum{(\rvx.)_{\zin-\Gamma}} (\mu)]
\;.
\eeq

Suppose $\cnet$ is the parent CB net of $\qnet$.
Suppose $\mu$ is the meta density matrix of  $\qnet$.
Then for any $\Gamma\subset \zn$, $P_\mu[(x.)_\Gamma]$ 
of $\qnet$ is identical to $P[(x.)_\Gamma]$ of $\cnet$. 
For example, if  $\qnet$ had nodes $\rva, \rvb, \rvc$ and amplitudes $A(a, b, c)$,
then $P_\mu(a,b,c)$ for $\qnet$ and $P(a, b, c)$ for $\cnet$
both equal $|A(a, b, c)|^2$. Likewise,
$P_\mu(a, b)$ for $\qnet$ and $P(a, b)$ for $\cnet$
both equal $\sum_c |A(a, b, c)|^2$.

We can define conditional probability distributions using 
the unconditional ones $P_\rho[ (x.)_{\Gamma}]$ defined above.
Suppose $\Gamma_1$ and $\Gamma_2$ are non-empty disjoint subsets of $\Gamma(\rho)$. Then
we define

\beq
P_\rho[ (x.)_{\Gamma_1}  |  (x.)_{\Gamma_2}]
=
\frac{
P_\rho[ (x.)_{\Gamma_1}, (x.)_{\Gamma_2}]
}{
\sum_{(y.)_{\Gamma_1}}P_\rho[ (y.)_{\Gamma_1}, (x.)_{\Gamma_2}]
}
=
\frac{
P_\rho[ (x.)_{\Gamma_1 \cup \Gamma_2}]
}{
P_\rho[ (x.)_{\Gamma_2}]
}
\;.
\label{eq:PA-def-cond-p-rho}\eeq
Note that

\beq
\sum_{(x.)_{\Gamma_1}}
P_\rho[ (x.)_{\Gamma_1}  |  (x.)_{\Gamma_2}]
=1
\;.
\eeq
Furthermore, 
if $\Gamma_1$, $\Gamma'_1$ and $\Gamma_2$
are non-empty disjoint subsets of $\Gamma(\rho)$, then 

\beq
\sum_{(x.)_{\Gamma'_1}}
P_\rho[ (x.)_{\Gamma_1 \cup \Gamma'_1}  |  (x.)_{\Gamma_2}]
=
P_\rho[ (x.)_{\Gamma_1}  |  (x.)_{\Gamma_2}]
\;.
\eeq

To illustrate the definition of $P[ (x.)_{\Gamma_1}  |  (x.)_{\Gamma_2}]$,
consider a density matrix $\rho$ which acts on $\hil_{\rva, \rvb, \rvc}$.
Then

\beq
P_\rho(a, b |  c) 
=
\frac{
P_\rho(a, b, c)
}{
\sum_{a',b'}
P_\rho(a', b' ,c)
}
=
\frac{
P_\rho(a, b, c)
}{
P_\rho(c)
}
\;,
\eeq	

\beq
\sum_{a,b} P_\rho(a,b  |  c) =1
\;,
\eeq

\beq
\sum_a P_\rho(a,b  |  c) = P_\rho(b  |  c)
\;.
\eeq

We can easily extend the definition Eq.(\ref{eq:PA-def-cond-p-rho})
of $P_\rho[ (x.)_{\Gamma_1}  |  (x.)_{\Gamma_2}]$
to the case that $\Gamma_1$ and $\Gamma_2$ overlap. 
We simply equate $P_\rho[ (x.)_{\Gamma_1}  |  (x.)_{\Gamma_2}]$ to 
$P_\rho[ (x.)_{\Gamma_1-\Gamma_2}  |  (x.)_{\Gamma_2}]$,
and evaluate the latter with definition Eq.(\ref{eq:PA-def-cond-p-rho}).

\EndSection


\BeginSection{Quantum Entropy: Its Definition and Properties}\label{sec:q-ent}
In this section, we will define various quantum entropies associated with
a QB net.

The von Neumann {\it quantum entropy} of a density matrix $\rho$ is defined by

\beq
S(\rho) = - \tr (\rho \log_2 \rho)
\;.
\label{eq:QE-s-def}\eeq

When $\rho$ is related to a QB net, it is convenient to
rephrase  Eq.(\ref{eq:QE-s-def}) in terms of the node random variables of the net. 
Consider a QB net  $\qnet$ with $N$ nodes
labelled by the random variables $\rvx_1, \rvx_2, \ldots, \rvx_N$.
Suppose $\rho$ is a density matrix 
that acts on the Hilbert space 
$\hil_{(\rvx.)_\Gamma(\rho)}$, and
suppose $\Gamma$, $\Gamma_1$ and $\Gamma_2 $ are non-empty subsets of $\Gamma(\rho)$.
$\Gamma_1$ and $\Gamma_2$ need not be disjoint.
We define:

\beq
S_\rho[ (\rvx.)_{\Gamma} ] = S[ \trace{ (\rvx.)_{\Gamma(\rho) - {\Gamma}} } (\rho)]
\;,
\label{eq:QE-s-simple}\eeq

\beq
S_\rho [ (\rvx.)_{\Gamma_1}  |  (\rvx.)_{\Gamma_2} ] =
S_\rho [ (\rvx.)_{\Gamma_1 \cup \Gamma_2} ] - S_\rho [ (\rvx.)_{\Gamma_2} ]
\;,
\label{eq:QE-s-cond}\eeq

\beq
S_\rho [ (\rvx.)_{\Gamma_1} : (\rvx.)_{\Gamma_2} ] =
S_\rho [ (\rvx.)_{\Gamma_1} ] +
S_\rho [ (\rvx.)_{\Gamma_2} ] -
S_\rho [ (\rvx.)_{\Gamma_1 \cup \Gamma_2} ] 
\;.
\label{eq:QE-s-mutual}\eeq
For example, suppose $\rva, \rvb, \rvc$ are nodes of a QB net.
If $\rho$ is a density matrix which acts on $\hila$, then

\beq
S_\rho(\rva) = S(\rho)
\;.
\eeq
If instead, $\rho$ acts on $\hil_{\rva, \rvb, \rvc}$, then

\beq
S_\rho(\rva) = S(\trace{\rvb, \rvc}\rho)
\;,
\eeq

\beq
S_\rho(\rva, \rvb) = S(\trace{ \rvc}\rho)
\;,
\eeq

\beq
S_\rho(\rva  |  \rvb) = S_\rho(\rva , \rvb) -S_\rho(\rvb)
\;,
\eeq

\beq
S_\rho(\rva : \rvb) = S_\rho(\rva) + S_\rho(\rvb) - S_\rho(\rva , \rvb)
\;.
\eeq
Eqs.(\ref{eq:QE-s-simple}) to (\ref{eq:QE-s-mutual}) for the quantum entropy $S_\rho(\cdot)$ 
are very natural generalizations 
of Eqs.(\ref{eq:CE-h-simple}) to (\ref{eq:CE-h-mutual}) for the classical entropy $H(\cdot)$.\cite{Cerf}

Note that definitions
Eqs.(\ref{eq:QE-s-simple}) to (\ref{eq:QE-s-mutual}) are independent of the order of the 
node random variables within $(\rvx.)_{\Gamma_1}$ and $(\rvx.)_{\Gamma_2}$.
For example, if $\rho$ is a density 
matrix acting on $\hil_{\rva, \rvb, \rvc}$, then

\beq
S_\rho(\rva, \rvb, \rvc) = S_\rho(\rva, \rvc, \rvb),
\;\;
S_\rho[\rva  |  (\rvb, \rvc)] = S_\rho[\rva  |  (\rvc, \rvb)]
\;.
\eeq
It is convenient to extend definitions Eqs.(\ref{eq:QE-s-simple}) to (\ref{eq:QE-s-mutual})
in the following two ways. First, we will allow
$(\rvx.)_{\Gamma_1}$ (ditto, $(\rvx.)_{\Gamma_2}$)
to contain repeated random variables. If it does, then 
we will throw out any extra copies of a random variable.
For example, if $\rho$ is a density 
matrix acting on $\hil_{\rva, \rvb, \rvc}$, then

\beq
S_\rho(\rva, \rva, \rvb, \rvc) = S_\rho(\rva, \rvb, \rvc),
\;\;
S_\rho[\rva  |  (\rvb, \rvb, \rvc)] = S_\rho[\rva  |  (\rvb, \rvc)]
\;.
\eeq
Second, we will allow
$(\rvx.)_{\Gamma_1}$ (ditto, $(\rvx.)_{\Gamma_2}$)
to contain internal parentheses. If it does, then 
we will ignore the internal  parentheses.
For example, if $\rho$ is a density 
matrix acting on $\hil_{\rva, \rvb, \rvc}$, then

\beq
S_\rho[(\rva, \rvb), \rvc] = S_\rho(\rva, \rvb, \rvc),
\;\; 
S_\rho[\rva  |  ((\rvb, \rvc), \rvd)] = S_\rho[\rva  |  (\rvb, \rvc, \rvd)]
\;.
\eeq

Let 
$\rvX = (\rvx.)_{\Gamma_1}$,
$\rvY = (\rvx.)_{\Gamma_2}$ and
$\rvZ = (\rvx.)_{\Gamma_3}$, where the
$\Gamma_1, \Gamma_2, \Gamma_3$ are
non-empty, possibly overlapping, subsets of $\zn$.
As with the function $H(\cdot)$, we will extend further the  domain of the function $S_\rho(\cdot)$ by introducing the
following axioms 

\beq
 S_\rho[ (\rvX, \rvY): \rvZ] = S_\rho[ (\rvX:\rvZ), (\rvY:\rvZ)]
\;,
\eeq

\beq
 S_\rho[ (\rvX: \rvY), \rvZ] = S_\rho[ (\rvX,\rvZ) : (\rvY,\rvZ)]
\;.
\eeq

Table \ref{Table-ent} gives a list of properties 
(identities and inequalities) satisfied by the quantum entropy $S_\rho(\cdot)$.
Whenever possible, Table \ref{Table-ent} matches each property of
the quantum entropy $S_\rho(\cdot)$ with an analogous property of
the classical entropy $H(\cdot)$.
 Analogous properties
are indicated by $H\rarrow S_\rho$. 
See Refs.\cite{Mans}-\cite{SchuLecs} to get  proofs of
those statements in Table \ref{Table-ent} that are not proven in this paper. 

An identity satisfied by $S(\cdot)$ but with no classical counterpart is:

\beq
S(U\rho U^\dagger) = S(\rho)
\;,
\label{eq:s-invariance}\eeq
for any unitary matrix $U$ acting on the same Hilbert space as the density matrix $\rho$.
We say that $S(\cdot)$ is invariant under unitary transformations of its argument.
Next we will 
rephrase Eq.(\ref{eq:s-invariance}) in terms of the node random variables of a QB net. 
Let $\Gamma_1$ and  $\Gamma_2$ be disjoint sets whose union is $\Gamma(\rho)$.
Define $X_1 = (x.)_{\Gamma_1}$, $X_2 = (x.)_{\Gamma_2}$, and $X = (x.)_{\Gamma(\rho)}$.
Thus, $X = (X_1, X_2)$.
$\rho$ acts on $\hil_{\rvX}$ so we can express it as:

\beq
\rho =
\sum_{ri}  \ket{ X}
\;
\rho_{X, X'}
\;
\bra{ X'}
\;.
\eeq
Suppose $U$ acts on $\hil_{\rvX_1}$. Then

\beq
\begin{array}{l}
U \rho U^\dagger =
\sum_{ri}
\ket{Y}\ket{X_2} 
U_{Y X_1}
\;
\rho_{(X_1, X_2),( X'_1, X'_2)}
\; 
U^\dagger_{X'_1 Y'}
\bra{X'_2}\bra{Y'}\\ 
\;\;=
\sum_{ri}
\ket{\psi_{\rvY}(X_1)}\ket{X_2}
\;
\rho_{(X_1, X_2),( X'_1, X'_2)}
\;
\bra{X'_2}\bra{\psi_{\rvY}(X'_1)}
\end{array}
\;,
\eeq
where

\beq
\ket{\psi_{\rvY}(X_1)} = \sum_{Y} \ket{\rvY = Y} U_{Y X_1}
\;.
\eeq
The Hilbert space $\hil_{\rvY}$ 
has the same dimension as $\hil_{\rvX_1}$.
The vectors $\ket{\psi_{\rvY}(X_1)}\in \hil_{\rvY}$
are orthonormal:

\beq
\av{\psi_{\rvY}(X_1)|\psi_{\rvY}(X'_1)} =
\delta(X_1, X'_1)
\;.
\eeq
Thus,

\beq
S_{U \rho U^\dagger}(\rvY, \rvX_2) 
=\left[S_{U \rho U^\dagger}(\rvY, \rvX_2)\right]_{U=1}
=S_\rho(\rvX_1, \rvX_2)
\;.
\eeq

Suppose $\rvX = (\rvx.)_{\Gamma}$ for some non-empty set $\Gamma\subset \Gamma(\rho)$.
The matrix $\trace{(\rvx.)_{\Gamma(\rho) - \Gamma}}(\rho)$
used in definition Eq.(\ref{eq:QE-s-simple}) of $S_\rho(\rvX)$ has diagonal entries which are
the probabilities $P_\rho(X)$ defined in Section \ref{sec:assoc-probs}. 
It is convenient to define a classical entropy for the $P_\rho(X)$
distribution:

\beq
H_\rho(\rvX) = - \sum_X P_\rho(X) \log_2 P_\rho(X)
\;.
\eeq
Because the probability distributions $P_\rho(X)$ are closed under marginalization,
$H_\rho(\cdot)$ satisfies all the identities and inequalities (see Table \ref{Table-ent}) satisfied by the
classical entropy $H(\cdot)$. 

It follows from Table \ref{Table-ent} that

\beq
0\leq S_\rho(\rvX) \leq H_\rho(\rvX)
\;.
\eeq
Thus, $H_\rho(\rvX)$ is a useful upper bound on $S_\rho(\rvX)$.
 
The quantities $H_\rho(\rvX)$ and 
$S_\rho(\rvX)$ complement each other 
in what they tells us about $\rho$ and $\rvX$. 
Indeed, note the following.
Suppose $\rvX = (\rvx.)_{\Gamma}$ where $\Gamma\subset \Gamma(\rho)$.
Let $\rho' = \trace{(\rvx.)_{\Gamma(\rho) - \Gamma}} (\rho)$ and
$M = \bra{(x.)_{\Gamma}} \rho' \ket{(x.)_{\Gamma}}$
so that  

\beq
S_\rho(\rvX)= -\tr(M \log_2 M)
\;,
\eeq

\beq
H_\rho(\rvX) = -\sum_i M_{ii} \log_2 M_{ii}
\;.
\eeq
$M$ is a diagonal matrix iff $S_\rho(\rvX)=H_\rho(\rvX)$.
Knowing $S_\rho(\rvX)$ alone does not tell us if $M$ is diagonal because
$\tr(M \log_2 M)$ is invariant under unitary transformations of $M$.

Henceforth, we will refer to the quantity 

\beq
Q_\rho(\rvX) = H_\rho(\rvX) - S_\rho(\rvX)
\;
\eeq
as the {\it coherence of $\rvX$ in $\rho$}. Note that

\beq
0\leq Q_\rho(\rvX) \leq \log_2 N_\rvX
\;.
\eeq
One has $Q_\rho(\rvX) = 0$ 
(i.e., zero coherence) iff $H_\rho(\rvX) = S_\rho(\rvX)$, which is true iff $M$ is diagonal. One has 
$Q_\rho(\rvX) = \log_2 N_\rvX$ (i.e., max. coherence) iff 
$S_\rho(\rvX) = 0$ and $H_\rho(\rvX) = \log_2 N_\rvX$.
$S_\rho(\rvX) = 0$ iff there exists some column vector $v$ such that
$M = v v^\dagger$. $H_\rho(\rvX) = \log_2 N_\rvX$ iff the diagonal entries of $M$ are all equal.
In fact, at max. coherence, all the entries of $M$ have the same absolute value $1/N_\rvX$.

$Q_\rho(\rvX)=0$ iff $\rho' = \trace{(\rvx.)_{\Gamma(\rho) - \Gamma}} (\rho)$
 is diagonal in the $\rvX$-basis $\{\ket{(x.)_{\Gamma}}\}$.
Hence, $Q_\rho(\rvX)$ can also be interpreted as the {\it mismatch between $\rho'$ and the $\rvX$ basis}. 
At zero mismatch, the $\rvX$ basis
constitutes a set of eigenvectors of $\rho'$.

\EndSection


\BeginSection{Mixed States and Purification}\label{sec:mixed-sts}

In this section, we will show how any mixed state density matrix can be 
represented by a QB net.

\begin{figure}[h]
	\begin{center}
	\epsfig{file=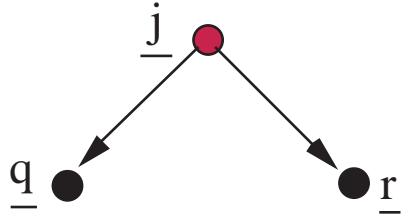}
	\caption{QB net for a mixed state.}
	\label{fig:mix-sts}
	\end{center}
\end{figure}

Consider the QB net of Fig.(\ref{fig:mix-sts}), where
\BeginBNetTabular
	$\rvj$  & $j = (j_1, j_2)$ & $\alpha_j$ & $\sum_j |\alpha_j|^2 = 1$\\
\hline
	$\rvq$ & $q$ & $\delta(q, j_1)$ & \\
\hline
	$\rvr$ & $r$ & $\delta(r, j_2)$ & \\
\EndBNetTabular

The meta density matrix $\mu$ for this net is 

\beq
\mu = \rhometa
\;,
\eeq 
where

\beq
\ketmeta = \sum_{ri} \alpha_{qr} \ket{ \rvj = (q, r)} \ket{ \rvq = q} \ket{\rvr = r}
\;.
\eeq
Define $\sigma$ and $\sigma_\rvq$ by

\beq
\sigma = \esum{\rvj}(\mu) = \sum_{ri} \alpha_{qr} \alpha^*_{q'r'} \ket{q,r}\bra{q',r'}
\;,
\eeq

\beq
\sigma_\rvq = \trace{\rvr}(\sigma) = \sum_{ri} \alpha_{qr} \alpha^*_{q'r} \ket{q}\bra{q'}
\;.
\eeq
Clearly, $\sigma$ is a pure state and $\sigma_\rvq$ is a mixed one. Since $\sigma$ 
is a pure state, 

\beq
S_\sigma(\rvq, \rvr) = 0
\;.
\eeq
By the Triangle Inequality (see Table \ref{Table-ent}),

\beq
S_\sigma(\rvq) = S_\sigma(\rvr)
\;.
\eeq

We've shown that some
mixed state density matrices can be represented by a QB net. But can any 
mixed state density matrix be represented in this manner? Yes.
This is why. Suppose $\rho$ is

\beq
\rho = \sum_{ri} \beta_{q q'} \ket{q}\bra{q'}
\;.
\eeq
Then the complex numbers $\beta_{q q'}$ define a Hermitian matrix $\beta$.
One can always decompose $\beta$ into $\beta = U \Gamma U^\dagger$,
where $U$ is a unitary matrix and $\Gamma$ is a diagonal matrix.
If we let $\alpha = U \sqrt{\Gamma}$, then

\beq 
\beta = \alpha \alpha^\dagger
\;. 
\eeq
Thus,

\beq
\rho =
\sum_{ri} 
\alpha_{qr} 
\alpha^*_{q'r}
\ket{q}\bra{q'}
\;.
\eeq
QED. The state

\beq
\ket{\psi} = \sum_{ri} \alpha_{qr}\ket{q, r}
\;
\eeq
is called a {\it purification} of $\rho$, because
the  mixed state $\rho$ can be obtained from the pure state $\ket{\psi}$
as follows: 

\beq
\trace{\rvr} \ket{\psi}\bra{\psi}
=\rho
\;.
\eeq

\EndSection


\BeginSection{Quantum System Interacting with Environment}\label{sec:sys+env}

In this section, we will consider QB nets that represents a quantum system interacting with its 
environment one or more times.

\subsection{Single Interaction}

\begin{figure}[h]
	\begin{center}
	\epsfig{file=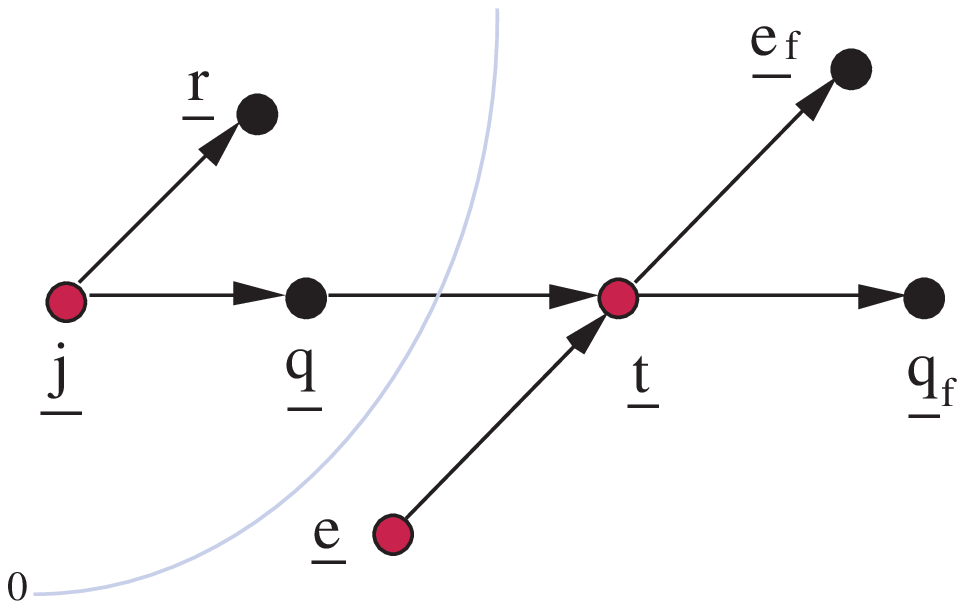}
	\caption{QB net for a system interacting once with its environment.}
	\label{fig:single}
	\end{center}
\end{figure}

Consider the QB net of Fig.(\ref{fig:single}), where
\BeginBNetTabular
	$\rvj$ 	& 	$j = (j^1,j^2)$ 	& 	$\alpha_j$  & $\sum_j |\alpha_j|^2 = 1$ \\
\hline
	$\rvq$	&	$q$					&	$\delta(q, j^1)$ &\\
\hline
	$\rvr$	&	$r$					&	$\delta(r, j^2)$ &\\
\hline
	$\rve$	&	$e$					&	$\beta_e$	&	$\sum_e |\beta_e|^2 = 1$	\\
\hline
	$\rvt$	&	$t=(t^1, t^2)$		&	$U(t |  q,e)$		& $\sum_t U(t  |  q,e) U^*(t  |  q',e') = \delta^{q'}_{q}\delta^{e'}_{e}$\\
\hline
	$\rvq_f$	&	$q_f$					&	$\delta(q_f, t^1)$	&\\
\hline
	$\rve_f$	&	$e_f$					&	$\delta(e_f, t^2)$	&\\
\EndBNetTabular

Let $\qnet$ be the QB net which contains all the nodes shown 
in Fig.(\ref{fig:single}). Let $ \qnet_0$ be the sub-net which contains only nodes $\rvj, \rvq_1, \rvr$.

The meta density matrix $\mu_0$ of $ \qnet_0$ is

\beq
\mu_0 = \ket{\psi_{meta}^0}\bra{\psi_{meta}^0}
\;,
\eeq
where

\beq
\ket{\psi_{meta}^0} = \sum_{ri} \alpha_{qr} \ket{\rvj = (q,r), q, r}
\;.
\eeq
Define $\rho_0$ by

\beq
\rho_0 = \esum{\rvj} \mu_0 = 
\sum_{ri} \alpha_{qr} \alpha_{q'r'}^* \ket{q,r}\bra{q',r'}
\;.
\eeq
$\rho_0$ is a pure state so 

\beq
S_{\rho_0}(\rvr, \rvq) = 0
\;.
\label{eq:SE-s-r-q}\eeq

The meta density matrix $\mu$ of $\qnet$ is

\beq
\mu = \rhometa
\;,
\eeq
where

\beq
\ketmeta = \sum_{ri} U(q_f, e_f  |  q, e) \beta_e \alpha_{qr} \ket{\rvj = (q,r), q, r, e, \rvt=(q_f, e_f), q_f, e_f}
\;.
\eeq
Define $\rho$ by

\beq
\rho = \esum{\rvj, \rvq, \rve, \rvt} \mu = 
\sum_{ri}
U(q_f, e_f |  q, e)\beta_e \alpha_{qr}
U^*(q'_f, e'_f |  q', e')\beta^*_{e'} \alpha^*_{q'r'}
\ket{r, q_f, e_f}\bra{r', q'_f, e'_f}
\;.
\eeq
$\rho$ is a pure state so 

\beq
S_{\rho}(\rvr, \rvq_f, \rve_f) = 0
\;.
\label{eq:SE-s-r-qf-ef}\eeq

By virtue of sub-additivity,

\beq
S_\rho(\rvr, \rve_f) - S_\rho(\rve_f) = S_\rho(\rvr  |  \rve_f) \leq S_\rho(\rvr)
\;.
\label{eq:SE-ineq-s-r-ef}\eeq
By Eqs.(\ref{eq:SE-s-r-q}) and (\ref{eq:SE-s-r-qf-ef}) and the Triangle Inequality,

\beq
S_\rho(\rvr, \rve_f) = S_\rho(\rvq_f)
\;,
\eeq

\beq
S_\rho(\rvr) = S_{\rho_0}(\rvr) = S_{\rho_0}(\rvq)
\;.
\eeq
Hence, Eq.(\ref{eq:SE-ineq-s-r-ef}) can be written as\cite{Schu96a}

\beq
S_\rho(\rvq_f) - S_\rho(\rve_f) \leq S_{\rho_0} (\rvq)
\;.
\eeq

\subsection{Multiple Interactions}

\begin{figure}[h]
	\begin{center}
	\epsfig{file=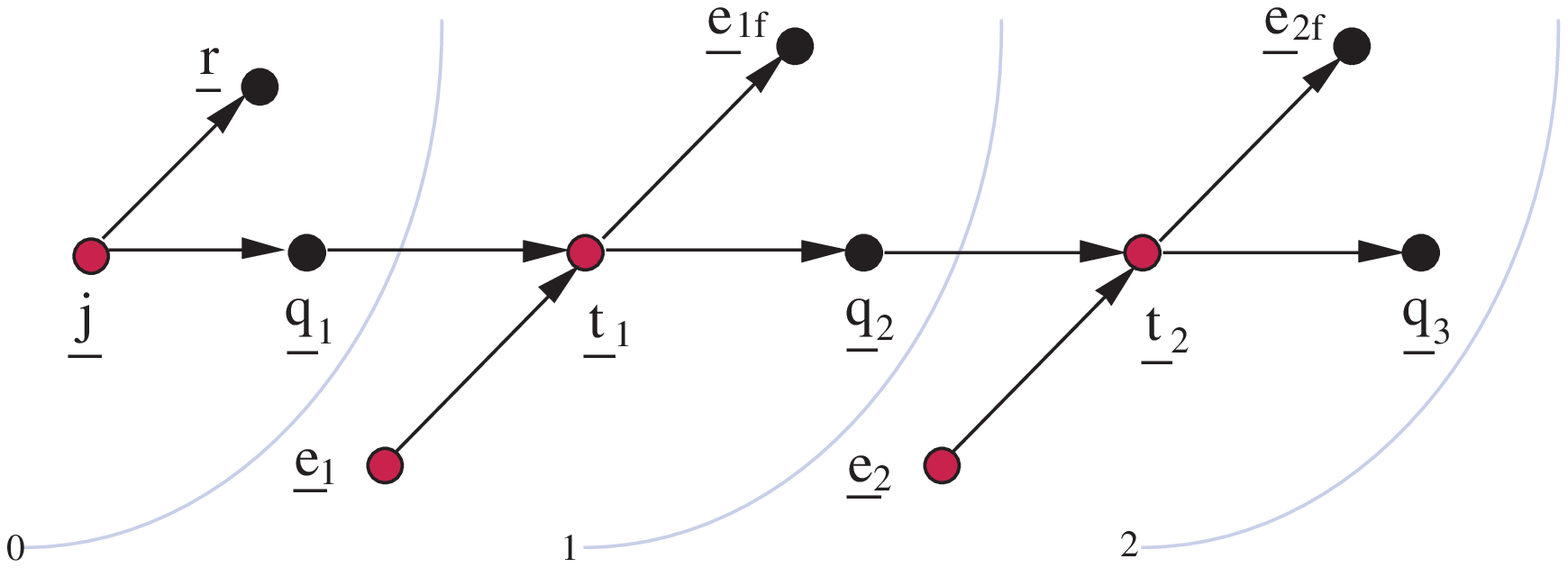, height=2.0in}
	\caption{QB net for a system interacting twice with its environment.}
	\label{fig:multi}
	\end{center}
\end{figure}

Consider the QB net of Fig.(\ref{fig:multi}), where
\BeginBNetTabular
	$\rvj$ 	& 	$j = (j^1,j^2)$ 	& 	$\alpha(j)$  & $\sum_j |\alpha(j)|^2 = 1$ \\
\hline
	$\rvq_1$	&	$q_1$			&	$\delta(q_1, j^1)$ &\\
\hline
	$\rvr$	&	$r$					&	$\delta(r, j^2)$ &\\
\hline
	$\rve_\lam$ for $\lam\in Z_{1,2}$	&	$e_\lam$					&	$\beta_\lam (e_\lam)$	&	$\sum_{e_\lam} |\beta_\lam(e_\lam)|^2 = 1$	\\
\hline
	$\rvt_\lam$ for $\lam\in Z_{1,2}$	&	$t_\lam=(t_\lam^1, t_\lam^2)$		&	$U_\lam(t_\lam |  q_\lam,e_\lam)$		& $\sum_{t_\lam}U_\lam(t_\lam  |  q_\lam,e_\lam) U^*_\lam(t_\lam  |  q'_\lam,e'_\lam) = \delta^{q'_\lam}_{q_\lam}\delta^{e'_\lam}_{e_\lam}$\\
\hline
	$\rvq_\lamf$ for $\lam\in Z_{1,2}$	&	$q_\lamf$					&	$\delta(q_\lamf, t^1_\lam)$	
	&Define $\rvq_2 = \rvq_{1f},\;\; \rvq_3 = \rvq_{2f}$\\
\hline
	$\rve_\lamf$ for $\lam\in Z_{1,2}$	&	$e_\lamf$					&	$\delta(e_\lamf, t^2_\lam)$	&\\
\EndBNetTabular

Let $ \qnet_0$ be the net which contains only nodes $\rvj, \rvq, \rvr$. For $\tau\in Z_{1, 2}$, let
$\qnet_\tau$ be the net which contains the previous net $\qnet_{\tau-1}$ plus nodes $\rve_\tau, \rvt_\tau, \rvq_{\tau f}, \rve_{\tau f}$.

For $\tau\in Z_{0, 2}$, the meta density matrix $\mu_\tau$ of net $\qnet_\tau$ is

\beq
\mu_\tau = \ket{\psi^\tau_{meta}} \bra{\psi^\tau_{meta}}
\;,
\eeq
where

\beq
\ket{\psi^\tau_{meta}} = \sum_{all} \left( \prod_{\lam=1}^\tau M_\lam\right)
\alpha(q_1, r)
\ket{ \rvj = (q_1,r), q_1, r}
\;,
\eeq
where

\beq
M_\lam = U_\lam( q_\lamf, e_\lamf  |  q_\lam, e_\lam) \beta_\lam(e_\lam) 
\ket{ e_\lam, \rvt_\lam = (q_\lamf, e_\lamf), q_\lamf, e_\lamf}
\;.
\eeq

Define $\rho_\tau$ for $\tau\in Z_{0, 2}$ by 

\beq
\rho_\tau = \esum{\rvX(\tau)} (\mu_\tau)
\;,
\eeq
where $\rvX(\tau)$ represents all the internal nodes of $\qnet_\tau$.
Thus,
$\rho_0$ acts on $\hil_{(\rvr, \rvq_1)}$,
$\rho_1$ acts on $\hil_{(\rvr, \rve_{1f},\rvq_2)}$, and
$\rho_2$ acts on $\hil_{(\rvr, \rve_{1f}, \rve_{2f},\rvq_3)}$.
For any $\tau\in Z_{0, 2}$, $\rho_\tau$ is a pure state	so

\begin{subequations}
\label{eq:SE-zero-ents}
\beq
S_{\rho_0}(\rvr, \rvq_1) = 0
\;,
\eeq

\beq
S_{\rho_1}(\rvr, \rve_{1f}, \rvq_{1f}) = 0
\;,
\eeq

\beq
S_{\rho_2}(\rvr, \rve_{1f}, \rve_{2f}, \rvq_{2f}) = 0
\;.
\eeq
\end{subequations}
Weak and strong sub-additivity imply 

\beq
S_{\rho_2}(\rvr  |  \rve_{1f},\rve_{2f})
\leq
S_{\rho_2}(\rvr  |  \rve_{1f} )
\leq
S_{\rho_2}(\rvr)
\;,
\eeq
which, by virtue of Eqs.(\ref{eq:SE-zero-ents}), can be written as\cite{Schu96a}

\beq
S_{\rho_2}(\rvq_{2f}) - S_{\rho_2}(\rve_{1f},\rve_{2f})
\leq
S_{\rho_1}(\rvq_{1f}) - S_{\rho_1}(\rve_{1f})
\leq
S_{\rho_0}(\rvq_{1})
\;.
\eeq

Define $\sigma_\tau$ for all $\tau\in Z_{0,2}$ by

\beq
\sigma_\tau = \esum{\rvX(\tau)} (\mu_\tau)
\;,
\eeq
where $\rvX(\tau)$ now represents all the internal nodes of 
$\qnet_\tau$ except for $\rvq_1, \rvq_2, \rvq_3$.
Thus,
$\sigma_0$ acts on $\hil_{(\rvr, \rvq_1)}$,
$\sigma_1$ acts on $\hil_{(\rvr, \rve_{1f},\rvq_1, \rvq_2)}$, and
$\sigma_2$ acts on $\hil_{(\rvr, \rve_{1f}, \rve_{2f},\rvq_1, \rvq_2, \rvq_3)}$.
Next we will show that

\beq
0 = S_{\sigma_0}(\rvq_1  |  \rvq_1) \leq
S_{\sigma_1}(\rvq_1  |  \rvq_2) \leq
S_{\sigma_2}(\rvq_1  |  \rvq_3)
\;,
\label{eq:SE-s-incr}\eeq
which is a quantum counterpart of the classical fixed sender DP inequality Eq.(\ref{eq:CBE-dp-ineq-cond}).
First note that

\begin{subequations}
\label{eq:SE-proof-s-incr}
\beq
S_{\sigma_2}(\rvq_1  |  \rvq_3) =
S_{\sigma_2}(\rvq_1  |  \rvq_{2f})\geq
S_{\sigma_2}(\rvq_1  |  \rvq_{2f}, \rve_{2f})
\;,
\eeq
where we've used $\rvq_3 = \rvq_{2f}$ and strong sub-additivity.
Since $S(\cdot)$ is invariant under unitary 
transformations of its argument, 

\beq
S_{\sigma_2}(\rvq_1  |  \rvq_{2f}, \rve_{2f}) =
\left[S_{\sigma_2}(\rvq_1  |  \rvq_{2f}, \rve_{2f})\right]_{U_2=1}=
S_{\sigma_1}(\rvq_1  |  \rvq_{2}) + nil
\;,
\eeq
where $nil$ equals $S(\sum_{ri} \beta_2(e_2)\beta^*(e'_2)\ket{e_2}\bra{e'_2})$,
which is zero. 
\end{subequations}
Combining Eqs.(\ref{eq:SE-proof-s-incr}), we get

\beq
S_{\sigma_2}(\rvq_1  |  \rvq_3)\geq 
S_{\sigma_1}(\rvq_1  |  \rvq_2)
\;.
\eeq
QED. For an alternative proof of Eq.(\ref{eq:SE-s-incr}), see \cite{DP_Ineq}.

\EndSection


\BeginSection{Two Mixtures Interacting}\label{sec:two-mixs}

In this section, we will consider a QB net that represents two mixed states scattering once off each other.

\begin{figure}[h]
	\begin{center}
	\epsfig{file=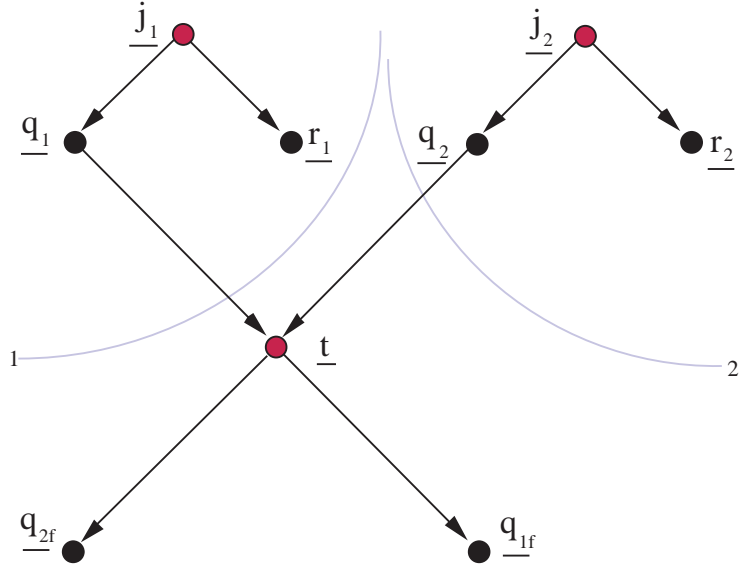, height=3in}
	\caption{QB net for 2 mixtures interacting.}
	\label{fig:scat}
	\end{center}
\end{figure}

Consider the QB net of Fig.(\ref{fig:scat}), where

\BeginBNetTabular
	$\rvj_\lam$ for $\lam\in Z_{1,2}$	& 	$j_\lam= (j_\lam^1,j_\lam^2)$ 
	& 	$\alpha_\lam(j_\lam)$  & $\sum_{j_\lam} |\alpha_\lam(j_\lam)|^2 = 1$ \\
\hline
	$\rvq_\lam$ for $\lam\in Z_{1,2}$	&	$q_\lam$			&	$\delta(q_\lam, j_\lam^1)$ &\\
\hline
	$\rvr_\lam$ for $\lam\in Z_{1,2}$	&	$r_\lam$			&	$\delta(r_\lam, j_\lam^2)$ &\\
\hline
	$\rvt$ &	$t =(t^1, t^2)$		&	$U(t |  q_1, q_2)$		& 
	$\sum_{t}U(t  |  q_1, q_2) U^*(t  |  q'_1, q'_2) = \delta^{q'_1}_{q_1}\delta^{q'_2}_{q_2}$\\
\hline
	$\rvq_\lamf$ for $\lam\in Z_{1,2}$	&	$q_\lamf$					&	$\delta(q_\lamf, t^\lam)$	&\\
\EndBNetTabular

Let $\qnet$ be the QB net which contains all the nodes shown 
in Fig.(\ref{fig:scat}). For $\lam\in Z_{1,2}$, let $\qnet_\lam$ be the sub-net which contains only nodes $\rvj_\lam, \rvq_\lam, \rvr_\lam$.

For $\lam\in Z_{1,2}$, the meta density matrix $\mu_\lam$ of $\qnet_\lam$ is

\beq
\mu_\lam = \ket{\psi_{meta}^\lam}\bra{\psi_{meta}^\lam}
\;,
\eeq
where

\beq
\ket{\psi_{meta}^\lam} = \sum_{ri} \alpha_\lam(q_\lam, r_\lam) \ket{\rvj_\lam = (q_\lam,r_\lam), q_\lam, r_\lam}
\;.
\eeq
Define $\rho_\lam$ by 

\beq
\rho_\lam = \esum{\rvj_\lam} \mu_\lam
\;.
\eeq
$\rho_\lam$ acts on $\hil_{(\rvq_\lam, \rvr_\lam)}$ and it is a pure state so 

\beq
S_{\rho_\lam}(\rvq_\lam, \rvr_\lam) = 0
\;.
\label{eq:TM-s-sub-net}\eeq

The meta density matrix $\mu$ of  $\qnet$ is

\beq
\mu = \rhometa
\;,
\eeq
where

\beq
\ketmeta = \sum_{ri} U(q_{1f}, q_{2f}  |  q_1, q_2) 
\left(
\prod_{\lam = 1}^2 \alpha_\lam(q_\lam, r_\lam) \ket{ \rvj_\lam = (q_\lam, r_\lam), q_\lam, r_\lam}
\right)
\ket{\rvt=(q_{1f},q_{2f}), q_{1f},q_{2f}}
\;.
\eeq
Define $\rho$ by:

\beq
\rho = \esum{\rvj_1, \rvq_1, \rvj_2, \rvq_2, \rvt} \mu
\;.
\eeq
$\rho$ acts on $\hil_{\rvr_1, \rvr_2, \rvq_{1f}, \rvq_{2f} }$ and it is a pure state so

\beq
S_\rho( \rvr_1, \rvr_2, \rvq_{1f}, \rvq_{2f} ) = 0
\;.
\label{eq:TM-s-full-net}\eeq

According to Table \ref{Table-ent},

\beq
| S_\rho( \rvq_\lamf) - S_\rho (\rvr_\lam) | 
\leq
S_\rho( \rvq_\lamf, \rvr_\lam) 
\leq
S_\rho( \rvq_\lamf) + S_\rho (\rvr_\lam)
\;,
\label{eq:TM-ineq}\eeq
for $\lam \in Z_{1,2}$.
By Eq.(\ref{eq:TM-s-sub-net}) and the Triangle Inequality,

\beq
S_\rho (\rvr_\lam) = S_{\rho_\lam} (\rvr_\lam) = S_{\rho_\lam} (\rvq_\lam)
\;.
\eeq
By Eq.(\ref{eq:TM-s-full-net}) and the Triangle Inequality,

\beq
S_\rho( \rvq_{1f}, \rvr_1) = S_\rho( \rvq_{2f}, \rvr_2)
\;.
\eeq
Hence, Eq.(\ref{eq:TM-ineq}) can be rewritten as\cite{Schu96b}

\beq
| S_\rho( \rvq_\lamf) - S_{\rho_\lam} (\rvq_\lam) | 
\leq
S_\rho( \rvq_{1f}, \rvr_1) = S_\rho( \rvq_{2f}, \rvr_2)
\leq
S_\rho( \rvq_\lamf) + S_{\rho_\lam} (\rvq_\lam)
\;.
\eeq
	
\EndSection


\BeginSection{POM}\label{sec:pom}
Given a Hilbert space $\hil_\rvq$, a {\it POM (Probability  Operator  Measure)}\cite{pom}
is a set $\{F_b  |  b \in S_\rvb \}$ of non-negative Hermitian operators
acting on $\hil_\rvq$. In addition, the observables $F_b$ must form a ``complete" set, meaning that

\beq
\sum_b F_b = 1
\;.
\label{eq:PO-complete}\eeq
If $\rho$ is a density matrix
acting on the same Hilbert space $\hil_\rvq$ as the $F_b$'s, then 
we can define a probability distribution for the
random variable $\rvb$ by

\beq
P(b) = \tr(\rho F_b)
\;,
\eeq
for all $b\in S_\rvb$.
We call an experiment that yields the value $b$ for $\rvb$ with a probability $P(b)$
a ``generalized measurement''.

We say that the $F_b$'s are (pairwise) orthogonal 
if $F_b F_{b'} = 0$ for all  $b,b' \in S_\rvb$ such that $b\neq b'$.
If the $F_b$'s are orthogonal, then we say that $\pom$ is 
an {\it orthogonal POM }.

An operator $F_b$ is said to have rank one if it can be represented in the form 
$\ket{\psi}\bra{\psi}$, where $\ket{\psi}$ need not have unit magnitude.
If $\ket{\psi}$ does have unit magnitude, then 
$F_b$ is a projector (i.e., $F_b^2 = F_b$).
An $F_b$ which is projector is a pure state density matrix.
For this reason, if the $F_b$'s are all projectors, then we say that $\pom$ is 
a {\it pure POM }.

A POM is both pure and orthogonal iff its $F_b$'s are (pairwise) orthogonal projectors
(i.e., $F_b F_{b'} = F_b\delta(b, b')$ for all  $b,b' \in S_\rvb$).
For such a POM,  we can represent each $F_b$ by $\ket{b}\bra{b}$, where 
the $\ket{b}$'s are an orthonormal basis of $\hil_\rvq$. Eq.(\ref{eq:PO-complete}) then reduces
to $\sum_b \ket{b}\bra{b} = 1$.
Such a POM is said to constitute a {\it von Neumann or ideal measurement}.

In this section, we will show how to represent 
a  POM as a QB net. Part (a)
will assume that
the $F_b$'s are orthogonal projectors.
Part (b) will not assume this.

\subsection{Orthogonal Projector $F_b$'s}\label{sec:pom-op}

\begin{figure}[h]
	\begin{center}
	\epsfig{file=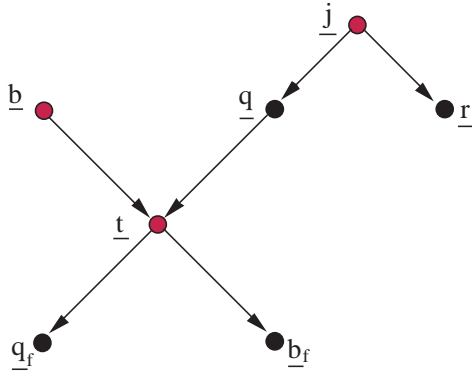, height=2in}
	\caption{QB net for orthogonal projector $F_b$'s.}
	\label{fig:ort}
	\end{center}
\end{figure}

Consider the QB net of Fig.(\ref{fig:ort}), where
\BeginBNetTabular
	$\rvj$  & $j = (j_1, j_2)$ & $\alpha_j$ & $\sum_j |\alpha_j|^2 = 1$\\
\hline
	$\rvq$ & $q$ & $\delta(q, j_1)$ & \\
\hline
	$\rvr$ & $r$ & $\delta(r, j_2)$ & \\
\hline
	$\rvt$ & $t = (t_1, t_2)$  & $U(t |  q, b)$  & $\sum_t U(t  |  q, b) U^*(t  |  q', b') = \delta^{q'}_q \delta^{b'}_b$\\
	&&& $U$ must also satisfy a constraint\\
	&&& equation relating it to the $F_b$'s.\\
\hline
	$\rvb$  & $b$ & $\delta(b, 0)$ & \\
\hline
	$\rvq_f$ & $q_f$ & $\delta(q_f, t_1)$ & \\
\hline
	$\rvb_f$ & $b_f$ & $\delta(b_f, t_2)$ & \\
\EndBNetTabular

Suppose the unitary operator $U$ satisfies:

\beq
U \ket{\phi}_\rvq \otimes \ket{0}_\rvb 
=
\sum_b 
\left( \sqrt{F_b} \;\ket{\phi}_\rvq \right )
\otimes
\ket{b}_\rvb
\;,
\label{eq:PO-a-constr-ops}\eeq
for any unit-magnitude vector $\ket{\phi}_\rvq \in \hil_{\rvq}$.
One can show that, for any POM $\pom$ acting on $\hil_\rvq$, there exists a 
unitary operator $U$ that
satisfies Eq.(\ref{eq:PO-a-constr-ops}).
Note that on the right-hand side of Eq.(\ref{eq:PO-a-constr-ops}), the state $\ket{b}$ acts as a pointer that
points towards a particular choice of $F_b$. 
Note that the completeness of the $F_b$'s and the unit-magnitude of $\ket{\phi}_\rvq$ 
together imply that
the right-hand side of Eq.(\ref{eq:PO-a-constr-ops}) is a 
unit-magnitude vector.
The vector $\ket{\phi}_\rvq \otimes \ket{0}_\rvb$ upon which
$U$ acts is likewise a unit-magnitude vector.
The fact that $U$ takes a unit-magnitude vector into another unit-magnitude vector 
(of the same dimension) is consistent
with the unitarity of $U$.

Eq.(\ref{eq:PO-a-constr-ops}) can be expressed in component form as follows:

\beq
\sum_{ri} U(q_f, b_f  |  q, b) \phi(q) \delta^b_0 = 
\sum_{ri} \sqrt{F_b} (q_f  |  q) \phi(q) \delta^{b_f}_b
\;
\label{eq:PO-a-constr-comps}\eeq
for any function $\phi(q)$. ($\phi(q)$ need not be normalized since it appears on both sides of the equation.)

Let $\qnet$ be the QB net which contains all the nodes shown 
in Fig.(\ref{fig:ort}). Let $\qnet_0$ be the sub-net which contains only nodes $\rvj, \rvq, \rvr$.

The meta density matrix $\mu_0$ of $\qnet_0$ is

\beq
\mu_0 = \ket{\psi_{meta}^0}\bra{\psi_{meta}^0}
\;,
\eeq
where

\beq
\ket{\psi_{meta}^0} = \sum_{ri} \alpha_{qr} \ket{\rvj = (q,r), q, r}
\;.
\eeq
Define 

\beq
\rho_0 = \esum{\rvj} \trace{\rvr}\mu_0 = 
\sum_{ri} \alpha_{qr} \alpha_{q'r}^* \ket{q}\bra{q'}
\;.
\eeq

The meta density matrix $\mu$ of  $\qnet$ is

\beq
\mu = \rhometa
\;,
\eeq
where 

\beq
\ketmeta = 
\sum_{ri} U(q_f, b_f  |  q, b) \alpha_{qr} \delta^b_0 
\ket{ \rvj =(q,r), q, r, b, \rvt = (q_f, b_f) , q_f, b_f}
\;.
\eeq
By Eq.(\ref{eq:PO-a-constr-comps}), $\ketmeta$ can also be expressed as

\beq
\ketmeta = 
\sum_{ri} \sqrt{F_b}(q_f  |  q) \alpha_{qr} \delta^{b_f}_b
\ket{ \rvj =(q,r), q, r, b, \rvt = (q_f, b_f) , q_f, b_f}
\;.
\eeq

Define $\rho$ by

\beq
\rho = \esum{\rvj, \rvq, \rvb, \rvt} \trace{\rvr, \rvq_f} \mu
\;.
\eeq
In other words, we get $\rho$ by tracing $\mu$ over all the external nodes except $\rvb_f$,
and e-summing it over all the internal nodes. $\rho$ acts on $\hil_{\rvb_f}$. 
Using the fact that the $F_b$'s are orthogonal projectors, it is easy to show that

\beq
\begin{array}{ll}
\rho  & = 
\sum_{ri} F_b(q'  |  q) \alpha_{qr} \alpha^*_{q'r}
\ket{\rvb_f = b} \bra{\rvb_f = b} \\
& = \sum_{ri} \tr (F_b \rho_0) \ket{\rvb_f = b} \bra{\rvb_f = b}
\end{array}
\;.
\eeq
Thus,

\beq
\av{b  |  \rho  |  b} = \tr (F_b \rho_0)
\;.
\eeq

\subsection{General $F_b$'s}\label{sec:pom-gen}

\begin{figure}[h]
	\begin{center}
	\epsfig{file=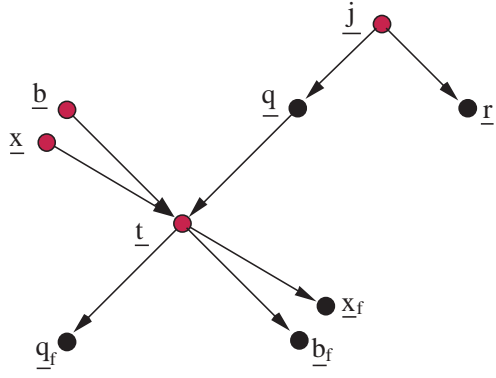, height=2in}
	\caption{QB net for general $F_b$'s.}
	\label{fig:non-ort}
	\end{center}
\end{figure}

Consider the QB net of Fig.(\ref{fig:non-ort}), where
\BeginBNetTabular
	$\rvj$  & $j = (j_1, j_2)$ & $\alpha_j$ & $\sum_j |\alpha_j|^2 = 1$\\
\hline
	$\rvq$ & $q$ & $\delta(q, j_1)$ & \\
\hline
	$\rvr$ & $r$ & $\delta(r, j_2)$ & \\
\hline
	$\rvt$ & $t = (t_1, t_2, t_3)$  & $U(t |  q, b, x)$  &
	$\sum_t U(t  |  q, b, x) U^*(t  |  q', b', x') = \delta^{q'}_q \delta^{b'}_b \delta^{x'}_x$\\
	&&& $U$ must also satisfy a constraint\\
	&&& equation relating it to the $F_b$'s.\\
\hline
	$\rvb$  & $b$ & $\delta(b, 0)$ & \\
\hline
	$\rvx$  & $x$ & $\delta(x, 0)$ & \\
\hline
	$\rvq_f$ & $q_f$ & $\delta(q_f, t_1)$ & \\
\hline
	$\rvb_f$ & $b_f$ & $\delta(b_f, t_2)$ & \\
\hline
	$\rvx_f$ & $x_f$ & $\delta(x_f, t_3)$ & \\
\EndBNetTabular

This is the same as the table in Section \ref{sec:pom-op}, except that there
are two new nodes ($\rvx, \rvx_f$), and the states of node $\rvt$ have
3 components instead of 2.

Instead of Eq.(\ref{eq:PO-a-constr-ops}), we now suppose the unitary operator $U$ satisfies:

\beq
U \ket{\phi}_\rvq \otimes \ket{0}_\rvb \otimes \ket{0}_\rvx
=
\sum_b 
\left( \sqrt{F_b} \;\ket{\phi}_\rvq \right )
\otimes
\ket{b}_\rvb
\otimes
\ket{b}_\rvx
\;,
\label{eq:PO-b-constr-ops}\eeq
for any unit-magnitude vector $\ket{\phi}_\rvq \in \hil_{\rvq}$.

Eq.(\ref{eq:PO-b-constr-ops}) can be expressed in component form as follows:

\beq
\sum_{ri} U(q_f, b_f, x_f |  q, b, x) \phi(q) \delta^b_0 \delta^x_0= 
\sum_{ri} \sqrt{F_b} (q_f  |  q) \phi(q) \delta^{b_f}_b \delta^{x_f}_b
\;.
\label{eq:PO-b-constr-comps}\eeq

Let $\qnet$ be the QB net which contains all the nodes shown 
in Fig.(\ref{fig:non-ort}). Let $\qnet_0$ be the sub-net which contains only nodes $\rvj, \rvq, \rvr$.

$\mu_0$ and $\rho_0$ are defined as in Section \ref{sec:pom-op} above.

The meta density matrix $\mu$ of  $\qnet$ is

\beq
\mu = \rhometa
\;,
\eeq
where 

\beq
\ketmeta = 
\sum_{ri} U(q_f, b_f, x_f  |  q, b, x) \alpha_{qr} \delta^b_0 \delta^x_0 
\ket{ \rvj =(q,r), q, r, b, x, \rvt = (q_f, b_f, x_f) , q_f, b_f, x_f}
\;.
\eeq
By Eq.(\ref{eq:PO-b-constr-comps}), $\ketmeta$ can also be expressed as

\beq
\ketmeta = 
\sum_{ri} \sqrt{F_b}(q_f  |  q) \alpha_{qr} \delta^{b_f}_b \delta^{x_f}_b
\ket{ \rvj =(q,r), q, r, b, x, \rvt = (q_f, b_f, x_f) , q_f, b_f, x_f}
\;.
\eeq

Define $\rho$ by

\beq
\rho = \esum{\rvj, \rvq, \rvb, \rvx, \rvt} \trace{\rvr, \rvq_f, \rvx_f} \mu
\;.
\eeq
In other words, we get $\rho$ by tracing $\mu$ over all the external nodes except $\rvb_f$,
and e-summing it over all the internal nodes. $\rho$ acts on $\hil_{\rvb_f}$. 
It is easy to show that

\beq
\rho  = \sum_{ri} \tr (F_b \rho_0) \ket{\rvb_f = b} \bra{\rvb_f = b}
\;.
\eeq
Thus,

\beq
\av{b  |  \rho  |  b} = \tr (F_b \rho_0)
\;.
\eeq
Whereas in Section \ref{sec:pom-op}, the orthogonal projector property of the $F_b$'s ``forces" $\rho$ to be diagonal, 
in this section, it is the tracing over node $\rvx_f$, a passive measurement of
that node, which forces $\rho$ to be diagonal.

\EndSection


\BeginSection{Signal Ensembles}\label{sec:signal-ens}
Suppose $\{ w_a  |  a\in \zn\}$ is a collection of
 non-negative numbers which add up to one. Suppose $\{ \rho_a  |  a\in \zn \}$
is a collection of density matrices all acting on the same Hilbert space $\hil$.
Let 
\beq
\rho = \sum_a w_a \rho_a
\label{eq:SE-rho}\;.
\eeq
We will say that $\rho$ is a {\it weighted sum of density matrices}.
We will call the collection $\cale = \enset$ a {\it signal ensemble}.
We will call the $w_a$'s the {\it weights} of $\cale$ and the $\rho_a$'s the {\it signal states} or {\it signals} of $\cale$.

In Quantum Information Theory, one is often interested in
density matrices like $\rho$ and ensembles like $\cale$.
One envisions sending a message encoded as a string
(for example: $\rho_1, \rho_5, \rho_3, \rho_1$) of  signal states.
(It is assumed that the states in the string are separated in some way, perhaps by intervening idle time periods.)
To say something about the average behavior of such messages,
one needs to consider $\rho$ and $\cale$.

We'll say two signals are {\it orthogonal} if 
$\rho_a \rho_b = 0$
for $a\neq b$. A signal ensemble such that all its signals are mutually orthogonal 
will be called an {\it orthogonal ensemble}.
Orthogonal ensembles play a special role in Quantum Information Theory, since their signals
are perfectly distinguishable (by a generalized measurement
with $F_b = \rho_b/(\sum_{b'} \rho_{b'})$. 
Suppose we are given a non-orthogonal signal ensemble $\enset$.
Then we can always replace it
by an orthogonal one. Indeed, 
if $\{ \ket{a} | a\in \zn\}$
is an orthonormal basis for some Hilbert space different from the one on which the
$\rho_a$'s act, and we define

\beq
\sigma_a = \ket{a}\bra{a} \rho_a
\;,
\eeq
for all $a$, then the ensemble $\cale' = \{ (w_a, \sigma_a) | \forall a\}$
is orthogonal.
Let

\beq
\sigma = \sum_a w_a \sigma_a = \sum_a w_a \ket{a}\bra{a} \rho_a
\;.
\label{eq:SE-sig}\eeq
Note how in $\sigma$, each projector $\ket{a}\bra{a}$ acts as
a pointer that points towards a particular choice of $\rho_a$.
We will say that $\rho$ of Eq.(\ref{eq:SE-rho}) (ditto,  $\sigma$ of Eq.(\ref{eq:SE-sig}) )
is a weighted sum of density matrices with {\it scalar weights}
(ditto, {\it orthogonal projector weights}).
Next, we will show how both
$\rho$ and $\sigma$ can be represented by a QB net.

\subsection{Scalar Weights}

\begin{figure}[h]
	\begin{center}
	\epsfig{file=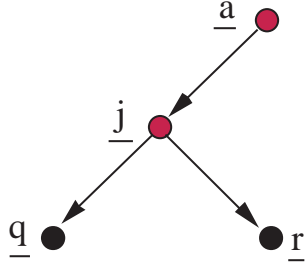, height=1.5in}
	\caption{QB net for a weighted sum of density matrices with scalar weights.}
	\label{fig:sca-wei}
	\end{center}
\end{figure}

Consider the QB net of Fig.(\ref{fig:sca-wei}), where
\BeginBNetTabular
	$\rva$ & $a$ & $\sqrt{w_a}$ & $\sum_a w_a = 1$ \\
\hline
	$\rvj$  & $j = (j_1, j_2)$ & $\alpha(j|a)$ & $\sum_j |\alpha(j|a)|^2 = 1$\\
\hline
	$\rvq$ & $q$ & $\delta(q, j_1)$ & \\
\hline
	$\rvr$ & $r$ & $\delta(r, j_2)$ & \\
\EndBNetTabular

The meta density matrix $\mu$ for this net is 

\beq
\mu = \rhometa
\;,
\eeq 
where

\beq
\ketmeta = \sum_{ri} \alpha(q,r|a) \sqrt{w_a}\;\ket{ a, \rvj = (q, r), q, r}
\;.
\eeq
If we define $\rho$ by

\beq
\rho = \esum{\rvj} \trace{\rva, \rvr} \mu
\;,
\eeq
then

\beq
\rho = \sum_a w_a \rho_a
\;,
\eeq
where

\beq
\rho_a = \sum_{ri/a} \alpha(q,r|a) \alpha^*(q',r|a) \ket{q}\bra{q'}
\;.
\eeq

\subsection{Orthogonal Projector Weights}

\begin{figure}[h]
	\begin{center}
	\epsfig{file=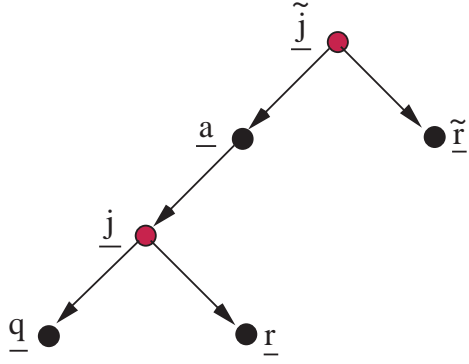, height=2in}
	\caption{QB net for a weighted sum of density matrices with orthogonal projector weights.}
	\label{fig:ort-pro}
	\end{center}
\end{figure}

Consider the QB net of Fig.(\ref{fig:ort-pro}), where
\BeginBNetTabular
	$\markrvj$  & $\markj = (\markj_1, \markj_2)$ & $\sqrt{w_{\markj_1}} \delta(\markj_1, \markj_2)$ & $\sum_{\markj_1} w_{\markj_1} = 1$\\
\hline
	$\rva$ & $a$ & $\delta(a, \markj_1)$ & \\
\hline
	$\markrvr$ & $\markr$ & $\delta(\markr, \markj_2)$ & \\
\hline
	$\rvj$  & $j = (j_1, j_2)$ & $\alpha(j|a)$ & $\sum_j |\alpha(j|a)|^2 = 1$\\
\hline
	$\rvq$ & $q$ & $\delta(q, j_1)$ & \\
\hline
	$\rvr$ & $r$ & $\delta(r, j_2)$ & \\
\EndBNetTabular

The meta density matrix $\mu$ for this net is 

\beq
\mu = \rhometa
\;,
\eeq 
where

\beq
\ketmeta = \sum_{ri} \alpha(q,r|a) \sqrt{w_a} \;
\ket{ \markrvj = (a, a), a, \markrvr=a }
\ket{ \rvj = (q, r), q, r}
\;.
\label{eq:SE-meta-ort-wei}\eeq

If we define 

\beq
\sigma = \esum{\markrvj, \rvj}\trace{\markrvr, \rvr}  \mu
\;,
\eeq
then

\beq
\sigma = \sum_a (w_a \ket{a}\bra{a} ) \rho_a
\;,
\eeq
where 

\beq
\rho_a = \sum_{ri/a} \alpha(q,r|a) \alpha^*(q',r|a) \ket{q}\bra{q'}
\;.
\eeq
Note that

\beq
S_\sigma(\rva, \rvq) = -\sum_a \trace{\rvq} \left [ w_a \rho_a \log_2 (w_a \rho_a) \right ]
= H(\vec{w}) + \sum_a w_a S(\rho_a)
\;,
\eeq

\beq
S_\sigma(\rvq) = S(\sum_a w_a \rho_a)
\;,
\eeq

\beq
S_\sigma(\rva) = H(\vec{w})
\;.
\eeq
Therefore,

\beq
S_\sigma(\rva : \rvq) = S(\sum_a w_a \rho_a) - \sum_a w_a S(\rho_a)
\;.
\eeq

\EndSection


\BeginSection{Signal Distinguishability}\label{sec:signal-dist}

In this section, we will define two measures
of signal distinguishability, the Holevo Information $\hol$ and
the Accessible Information $\acc$. Then we will 
use a QB net to prove 
that $\acc\leq\hol$, a result know as Holevo's Inequality\cite{Hol73}.

\subsection{Holevo Information}

Given a signal ensemble $\cale = \enset$, let 

\beq
\rho = \sum_a w_a \rho_a
\;.
\eeq
 The
{\it Holevo Information} is defined by

\beq
\hol = S(\rho) - \sum_a w_a S(\rho_a)
\;.
\eeq

Some of the properties of $\hol$ are:
\begin{enumerate}

\item[(a)]
If the $\rho_a$'s are pure states, then
$\hol = S(\rho)$.

\item[(b)]
If the $\rho_a$'s are all the same, then $\hol = 0$.
This result can be generalized as follows.
The convexity of $S(\cdot)$ (see Table \ref{Table-ent}) implies 
$0 \leq \hol$, 
with equality iff the $\rho_a$'s are all the same. Thus, $\hol$ 
measures the indistinguishability of the signal states.

\item[(c)]
If the $\rho_a$'s are orthogonal, then

\beq
S(\rho) =
-\sum_a \tr [ w_a \rho_a \log_2 ( w_a \rho_a ) ] 
\;,
\label{eq:SD-collapse}\eeq
because orthogonal  $\rho_a$'s ``don't mix" with each other
so  all sums over index $a$ collapse into a single outside sum.
From Eq.(\ref{eq:SD-collapse}), it follows that

\beq
S(\rho) =  
-\sum_a \tr \left[ w_a \rho_a (\log_2  w_a + \log_2 \rho_a) \right]
= H(\vec{w}) + \sum_a w_a S(\rho_a)
\;,
\eeq
so 

\beq
\hol = H(\vec{w})
\;.
\eeq
We see that since orthogonal states are completely distinguishable, 
their quantum entropy is essentially classical.
This result can be generalized as follows.
According to Table \ref{Table-ent}, 

\beq
\hol \leq H(\vec{w})
\;,
\eeq
with equality iff the $\rho_a$'s are orthogonal.

\item[(d)]
If the $\rho_a$'s commute (i.e., $\rho_a \rho_b = \rho_b \rho_a$ for all $a, b$), then 
$\hol$ reduces to a classical entropy. Indeed, because of the commutativity, the $\rho_a$'s can be 
simultaneously  diagonalized in an orthonormal basis $\{ \ket{b} | \forall b\}$. In this basis, 
$S(\rho_a)$ for all $a$ and $S(\rho)$ reduce to classical entropies. To calculate $\hol$ explicitly, 
define probabilities $P(a|b)$ and $P(a)$ by

\beq
\rho_a = \sum_b P(b|a) \ket{b}\bra{b}
\;,
\eeq

\beq
P(a) = w_a
\;.
\eeq
Then 

\beq
S(\rho) = S(\sum_{a,b} P(a,b) \ket{b}\bra{b})
= S(\sum_b P(b) \ket{b}\bra{b}) = H(\rvb)
\;,
\eeq

\beq
\sum_a w_a S(\rho_a) =  \sum_a P(a) \{-\sum_b P(b|a) \log_2 P(b|a) \}= H(\rvb | \rva)
\;,
\eeq
so

\beq
\hol = H(\rva : \rvb)
\;.
\eeq

\end{enumerate}

\subsection{Accessible Information}

Suppose Alice sends Bob a signal $\rho_{a_0}$ using the signal ensemble $\cale = \enset$.
Bob knows which ensemble Alice is using, but he doesn't know $a_0$. To guess 
$a_0$, Bob devises and measures a POM $\pom$. The value $b$ that he measures
for $\rvb$ will 
be characterized by:

\beq
P(b|a) = \tr( F_b \rho_a)
\;.
\label{eq:SD-cond}\eeq
(This probability distribution specifies a so called {\it quantum channel}.)
Since Bob knows $\cale$, he can use 

\beq
P(a) = w_a
\;
\label{eq:SD-prior}\eeq
as the a priori probability for signal $\rho_a$ for all $a\in S_\rva$. 
Bob would like to determine the posterior probabilities
$P(a|b)$ in terms of what he knows ($P(b|a)$ and $P(a)$). He can do this with Bayes'
rule:

\beq
P(a|b) = \frac{P(b|a) P(a)}{\sum_{a'} P(b|a') P(a')}
\;.
\eeq
Bob will guess $a_0$ best if  he uses the  magical POM $\pom$ that 
minimizes the $a$ spread of the probability distribution $P(a|b)$. This spread is measured by $H(\rva | \rvb)$. But
$H(\rva : \rvb)$ (called the ``transmitted information")
 equals  $H(\rva) - H(\rva | \rvb)$ and $H(\rva)$ is $F_b$ independent. 
So the magical POM also
maximizes the transmitted information $H(\rva : \rvb)$.

For any signal ensemble $\cale = \enset$, we define
the {\it Accessible Information} by

\beq
\acc = \max_{\pom } H(\rva : \rvb)
\;,
\eeq
where $P(b|a)$ and $P(a)$ are defined by Eqs.(\ref{eq:SD-cond}) and (\ref{eq:SD-prior}).
Since mutual entropies are always non-negative, $\acc\geq 0$.
One can show that equality is achieved iff the $\rho_a$'s are all the same.
Hence, $\acc$ is a measure of indistinguishability of the signals $\rho_a$, just 
like $\hol$ is. In fact, these two measures of indistinguishability
are related by the so called Holevo's Inequality\cite{Hol73}:

\beq
\acc \leq \hol
\;,
\eeq
which we will prove in the next section.
It makes intuitive sense that $\acc$ is both a measure of 
indistinguishability and a measure of maximum information transmission.
One expects that making more distinguishable the signals which compose a message will 
increase the information transmitted by the message.

\subsection{Holevo's Inequality}

Next, we will use a QB net to prove Holevo's Inequality.

\begin{figure}[h]
	\begin{center}
	\epsfig{file=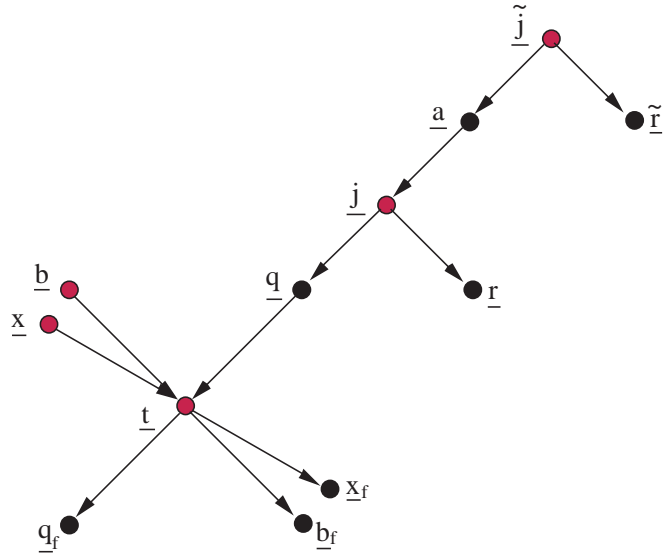, height=3in}
	\caption{QB net for proving Holevo's Inequality.}
	\label{fig:holevo}
	\end{center}
\end{figure}

Consider the QB net of Fig.(\ref{fig:holevo}), where
\BeginBNetTabular
	$\markrvj$  & $\markj = (\markj_1, \markj_2)$ & $\sqrt{w_{\markj_1}}\;\delta(\markj_1, \markj_2)$ 
	& $\sum_{\markj_1} w_{\markj_1} = 1$\\
\hline
	$\rva$ & $a$ & $\delta(a, \markj_1)$ & \\
\hline
	$\markrvr$ & $\markr$ & $\delta(\markr, \markj_2)$ & \\
\hline
	$\rvj$  & $j = (j_1, j_2)$ & $\alpha(j|a)$ & $\sum_j |\alpha(j|a)|^2 = 1$\\
\hline
	$\rvq$ & $q$ & $\delta(q, j_1)$ & \\
\hline
	$\rvr$ & $r$ & $\delta(r, j_2)$ & \\
\hline
	$\rvt$ & $t = (t_1, t_2, t_3)$  & $U(t |  q, b, x)$  &
	$\sum_t U(t  |  q, b, x) U^*(t  |  q', b', x') = \delta^{q'}_q \delta^{b'}_b \delta^{x'}_x$\\
	&&& $U$ must also satisfy a constraint\\
	&&& equation relating it to the $F_b$'s.\\
\hline
	$\rvb$  & $b$ & $\delta(b, 0)$ & \\
\hline
	$\rvx$  & $x$ & $\delta(x, 0)$ & \\
\hline
	$\rvq_f$ & $q_f$ & $\delta(q_f, t_1)$ & \\
\hline
	$\rvb_f$ & $b_f$ & $\delta(b_f, t_2)$ & \\
\hline
	$\rvx_f$ & $x_f$ & $\delta(x_f, t_3)$ & \\
\EndBNetTabular

The matrix $U$ must implement a general POM $\pom$. Hence, it will be assumed to satisfy
Eq.(\ref{eq:PO-b-constr-comps}), which we restate:

\beq
\sum_{ri} U(q_f, b_f, x_f |  q, b, x) \phi(q) \delta^b_0 \delta^x_0= 
\sum_{ri} \sqrt{F_b} (q_f  |  q) \phi(q) \delta^{b_f}_b \delta^{x_f}_b
\;,
\label{eq:SD-u-fb}\eeq
for any function $\phi(q)$.

Let $\qnet_f$ be the QB net which contains all the nodes shown 
in Fig.(\ref{fig:holevo}). Let $ \qnet_0$ be the sub-net which contains only nodes 
$\markrvj, \rva, \markrvr, \rvj, \rvq, \rvr$.

The meta density matrix $\mu^0$ of $\qnet_0$ was specified in  Eq.(\ref{eq:SE-meta-ort-wei}).
We also showed in Section \ref{sec:signal-ens} that if $\rho^0$ is defined by

\beq
\rho^0 = \esum{\markrvj, \rvj}\trace{\markrvr, \rvr}  \mu^0
\;,
\eeq
then

\beq
\rho^0 = \sum_a (w_a \ket{a}\bra{a} ) \rho_a
\;,
\eeq
where 

\beq
\rho_a = \sum_{ri/a} \alpha(q,r|a) \alpha^*(q',r|a) \ket{q}\bra{q'}
\;.
\eeq
Furthermore, we showed that if $\cale = \enset$, then 

\beq
S_{\rho^0}(\rva : \rvq) = \hol
\;.
\label{eq:SD-chi}\eeq

The meta density matrix $\mu^f$ of $\qnet_f$ is 

\beq
\mu^f = \ket{\psi_{meta}^f}\bra{\psi_{meta}^f}
\;,
\eeq
where

\beq
\begin{array}{l}
\ket{\psi_{meta}^f} = \sum_{ri} 
U(q_f, b_f, x_f | q, b, x)\delta^b_0 \delta^x_0
\alpha(q,r|a) \sqrt{w_a}\\
\;\;\;
\ket{ \markrvj = (a, a), a, \markrvr=a ,\rvj = (q, r), q, r, b, x, \rvt=(q,b,x), q_f, b_f, x_f}
\end{array}
\;.
\eeq
Define $\rho^f$ by

\beq
\rho^f = \trace{\rvr, \markrvr} \esum{\markrvj, \rvj, \rvq, \rvb, \rvx, \rvt}(\mu^f)
\;.
\eeq
In other words, we trace $\mu^f$ over all the external nodes 
except $q_f, b_f, x_f$, and we e-sum it over all internal ones except $\rva$.
Hence, $\rho^f$ acts on $\hil_{ \rvq_f, \rvb_f, \rvx_f, \rva}$. 

To prove Holevo's Inequality, we begin by noticing that

\begin{subequations}
\label{eq:SD-rhof-to-rhoo}
\beq
S_{\rho^f}(\rva : (\rvb_f, \rvq_f, \rvx_f) ) =
S_{\rho^f}(\rva)
+S_{\rho^f}(\rvb_f, \rvq_f, \rvx_f)
-S_{\rho^f}(\rva , \rvb_f, \rvq_f, \rvx_f)
\;,
\eeq

\beq
S_{\rho^f}(\rva) = S_{\rho^0}(\rva)
\;,
\eeq

\beq
S_{\rho^f}(\rvb_f, \rvq_f, \rvx_f) = \left[S_{\rho^f}(\rvb_f, \rvq_f, \rvx_f)\right]_{U=1} = S_{\rho^0}(\rvq)
\;,
\eeq

\beq
S_{\rho^f}(\rva, \rvb_f, \rvq_f, \rvx_f) = \left[S_{\rho^f}(\rva, \rvb_f, \rvq_f, \rvx_f)\right]_{U=1} = S_{\rho^0}(\rva, \rvq)
\;.
\eeq
\end{subequations}
Combining Eqs.(\ref{eq:SD-rhof-to-rhoo}) yields

\begin{subequations}
\label{eq:SD-h-to-chi}
\beq
S_{\rho^f}[\rva : (\rvb_f, \rvq_f, \rvx_f) ] =
S_{\rho^0}(\rva : \rvq)
\;.
\eeq
By virtue of strong sub-additivity,

\beq
S_{\rho^f}(\rva : \rvb_f ) 
\leq
S_{\rho^f}[\rva : (\rvb_f, \rvq_f, \rvx_f) ] 
\;.
\eeq
Below, we will show that

\beq
H_{\rho^f}(\rva : \rvb_f ) =
S_{\rho^f}(\rva : \rvb_f ) 
\;.
\label{eq:SD-postponed}\eeq
\end{subequations}
Combining Eqs.(\ref{eq:SD-chi}) and (\ref{eq:SD-h-to-chi}) finally yields
Holevo's Inequality:

\beq
H_{\rho^f}(\rva : \rvb_f ) \leq
S_{\rho^0}(\rva : \rvq) = \hol
\;.
\eeq
This can be understood as a special case of the Fixed Sender Data Processing Inequality
\cite{DP_Ineq},\cite{Winter}. It 
says that when information is transmitted from $\rva$, 
less reaches  $\rvb_f$ than $\rvq$.

To show Eq.(\ref{eq:SD-postponed}), we use Eq.(\ref{eq:SD-u-fb})
to express $\rho^f$ in terms of the POM $\pom$.
It is then easy to show that

\beq
\trace{\rvq_f, \rvx_f} \rho^f =
\sum_{ri}
\tr(F_b \rho_a) w_a \ket{a,b}\bra{a,b}
\;.
\eeq
Replacing
$\tr(F_b \rho_a)$
and 
$w_a$
by
$P(b|a)$ and $P(a)$ (see Eqs.(\ref{eq:SD-cond}) and (\ref{eq:SD-prior}))
yields

\beq
\trace{\rvq_f, \rvx_f} \rho^f = 
\sum_{a,b}
P(a,b)
 \ket{a,b}\bra{a,b}
\;.
\eeq
Eq.(\ref{eq:SD-postponed}) now follows. 

\subsection{Example}

\begin{figure}[h]
	\begin{center}
	\epsfig{file=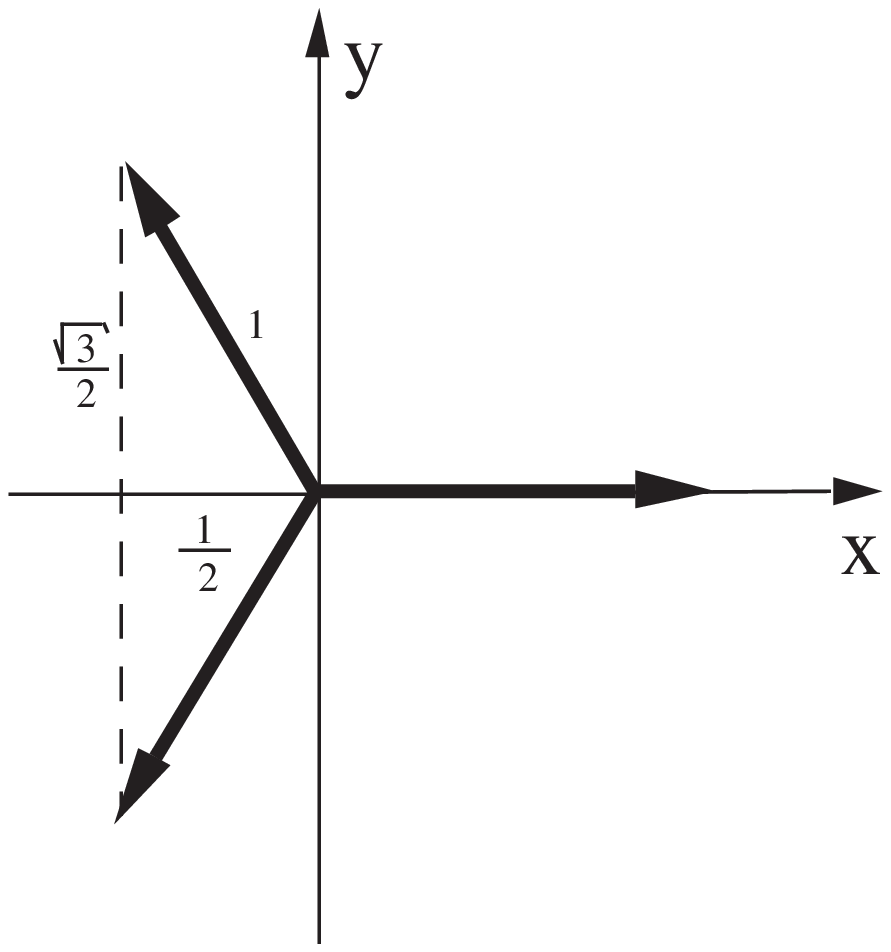, height=2.5in}
	\caption{The vectors $\ket{\phi_1}$, $\ket{\phi_2}$, $\ket{\phi_3}$.}
	\label{fig:3-vecs}
	\end{center}
\end{figure}

The following example (originally from Ref.\cite{PeWo}) is often used to illustrate Holevo's  Bound.

Let

\beq
\ket{\phi_1}=
\left[
\begin{array}{c}
1\\
0
\end{array}
\right]
\;,
\;\;
\ket{\phi_2}=
\frac{1}{2}
\left[
\begin{array}{c}
-1\\
\sqrt{3}
\end{array}
\right]
\;,
\;\;
\ket{\phi_3}=
\frac{1}{2}
\left[
\begin{array}{c}
-1\\
-\sqrt{3}
\end{array}
\right]
\;.
\label{eq:SD-phis}\eeq
As shown in Fig.(\ref{fig:3-vecs}), these 3 vectors specify the corners of an equilateral 
triangle that lies on the real plane. Now consider the signal ensemble $\cale = \enset$, with

\beq
w_a = \frac{1}{3}
\;,
\eeq

\beq
\rho_a = \ket{\phi_a}\bra{\phi_a}
\;,
\eeq
for $a\in Z_{1,3}$. It is easy to show that

\beq
\rho = \sum_{a=1}^3 w_a \rho_a = 
\frac{1}{2}
\left[ 
\begin{array}{cc}
1&0 \\
0&1
\end{array}
\right]
\;,
\eeq
so

\beq
\hol = S(\rho) = 1
\;.
\eeq
Define a POM $\pom$ by

\beq
F_b = \frac{2}{3} (1 - \ket{\phi_b}\bra{\phi_b})
\;,
\eeq
where $b \in Z_{1,3}$. Then

\beq
P(b|a) = \av{ \phi_a | F_b | \phi_a} =
\left\{
\begin{array}{cc}
0\;\;{\rm if}\;\; a=b\\
\frac{1}{2}\;\;{\rm if}\;\; a\neq b
\end{array}
\right.
\;.
\eeq
According to Bayes' rule, in this case
the posterior probabilities $P(a|b)$ are equal to $P(b|a)$.
Thus, if Bob measures this POM and obtains the value $b$, he can safely conclude 
that Alice did not send signal $b$, 
and he can assign equal posterior probabilities to the other two signals. One can show that
this POM maximizes $H(\rva : \rvb)$. Therefore,

\beq
\acc = H(\rva : \rvb) = .5850
\;.
\eeq
Holevo's Inequality is satisfied, as expected.

Another interesting ensemble considered in Refs.\cite{PeWo} and \cite{PresBook} is

\beq
w_a = \frac{1}{3}
\;,
\eeq

\beq
\rho_a = \ket{\Phi_a}\bra{\Phi_a}
\;,
\eeq

\beq
\ket{\Phi_a} = \ket{\phi_a} \otimes \ket{\phi_a}
\;,
\eeq
where $a\in Z_{1,3}$, and the vectors $\ket{\phi_a}$ are those defined previously in Eq.(\ref{eq:SD-phis}).
One finds $\hol = 1.5$ and $\acc = 1.3691$.

\EndSection


\BeginSection{EPR Pair}\label{sec:EPR-pair}

In this section, we will consider a QB net that represents an EPR pair.
An EPR pair consists of two spin half particles in  a singlet state  (i.e., a state of zero total spin).

Suppose $\ket{+_z}$ and  $\ket{-_z}$ are the states of spin up and down in the +Z direction.
We define $\ket{\psi_{EPR}}$ by

\beq
\ket{\psi_{EPR}} 
=
\frac{1}{\sqrt{2}}
( \ket{+_z}\otimes \ket{-_z} -
\ket{-_z}\otimes \ket{+_z} )
\;.
\eeq
Let

\beq
\ket{+_z} = \ket{0} = 
\left [
\begin{array}{l}
1 \\
0
\end{array}
\right ]
\;,
\eeq

\beq
\ket{-_z} = \ket{1} = 
\left [
\begin{array}{l}
0 \\
1
\end{array}
\right ]
\;.
\eeq
If $e=(e_1, e_2)\in Bool^2$, then $\psi_{EPR}(e) = \av{e | \psi_{EPR}}$ is

\beq
\psi_{EPR}(e) = \frac{1}{\sqrt{2}}[ \delta^{e_1,e_2}_{0,1} - \delta^{e_1,e_2}_{1,0}]
\;.
\eeq

\begin{figure}[h]
	\begin{center}
	\epsfig{file=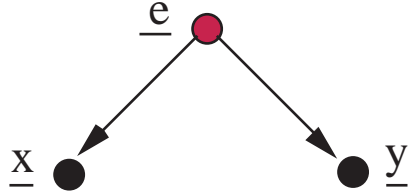}
	\caption{QB net for EPR pair.}
	\label{fig:pair}
	\end{center}
\end{figure}

Consider the QB net of Fig.(\ref{fig:pair}), where
\BeginBNetTabular
	$\rve$ & $e = (e_1, e_2) \in Bool^2$ & $\psi_{EPR}(e) = \frac{1}{\sqrt{2}}[ \delta^{e_1,e_2}_{0,1} - \delta^{e_1,e_2}_{1,0}]$ & \\
\hline
	$\rvx$ & $x\in Bool$ & $\delta(x, e_1)$ & \\
\hline
	$\rvy$ & $y\in Bool$ & $\delta(y, e_2)$ & \\
\EndBNetTabular

The meta density matrix $\mu$ of this net is

\beq
\mu = \rhometa
\;,
\eeq
where

\beq
\ketmeta = \sum_{ri} \psi_{EPR}(x, y) \ket{ \rve = (x, y), x, y}
\;.
\eeq
Define $\rho$ by:

\beq
\rho = \esum{\rve} \mu
\;.
\eeq
Then

\beq
\rho = \sum_{ri} \frac{1}{2} ( \delta^{x,y}_{0,1} - \delta^{x,y}_{1,0})
( \delta^{x',y'}_{0,1} - \delta^{x',y'}_{1,0})
\ket{x, y} \bra{x', y'}
\;,
\eeq

\beq
[\bra{x,y} \rho \ket{x', y'}] = 
\frac{1}{2}
\left[
\begin{array}{r}
0\\ 1\\ -1\\ 0
\end{array}
\right ]
\left[
\begin{array}{rrrr}
0 & 1 & -1 & 0
\end{array}
\right ]
=
\frac{1}{2}
\begin{tabular}{r|rrrr}
         & {\tiny 00} & {\tiny 01} & {\tiny 10} & {\tiny 11}\\
\hline
{\tiny 00}&  0&  0&  0&  0\\
{\tiny 01}&  0& 1&  -1& 0\\
{\tiny 10}&  0&  -1& 1& 0\\
{\tiny 11}&  0& 0& 0&  0\\
\end{tabular}
\;.
\eeq
$\rho$ is a pure state so $S_\rho(\rvx, \rvy) = 0$ and $S_\rho(\rvx) = S_\rho(\rvy)$.
It is easy to show that

\beq
\tr_\rvx \rho = \frac{1}{2} \sum_y \ket{y}\bra{y}
\;,
\eeq

\beq
\tr_\rvy \rho = \frac{1}{2} \sum_x \ket{x}\bra{x}
\;,
\eeq
Thus,

\beq
\begin{tabular}{lll}
$S_\rho(\rvx) = 1,$ 			& $H_\rho(\rvx) = 1$			&	(zero coherence) \\
$S_\rho(\rvy) = 1,$				& $H_\rho(\rvy) = 1$			&	(zero coherence) \\
$S_\rho(\rvx, \rvy) = 0,$ 		& $H_\rho(\rvx, \rvy) = 1$		&	(not max. coherence since $H_\rho(\rvx, \rvy) \neq 2$) \\
\\
$S_\rho(\rvx |  \rvy) = -1,$ 	& $H_\rho(\rvx |  \rvy) = 0$ &\\
$S_\rho(\rvy |  \rvx) = -1,$	& $H_\rho(\rvy |  \rvx) = 0$ &\\
$S_\rho(\rvx: \rvy) =2,$ 		& $H_\rho(\rvx: \rvy) = 1$ &\\
\end{tabular}
\;.
\label{eq:EPR-ents-rho}\eeq

Define $\rho(y)$ by

\beq
\rho(y) = \esum{\rve}\bra{y} \mu \ket{y} = \bra{y} \rho \ket{y}
\;.
\eeq
$\rho(y)$ acts on $\hil_\rvx$. It is easy to show that

\beq
\rho(y) = \ket{0}\bra{0} \delta^y_1 + \ket{1}\bra{1} \delta^y_0 
\;.
\eeq
Thus,

\beq
\begin{tabular}{lll}
$S_{\rho(y)}(\rvx) = 0,$ 			& $H_{\rho(y)}(\rvx) = 0$			&	(zero coherence)
\end{tabular}
\;.
\eeq

These results can be interpreted as follows. We start with an EPR pair of particles. One 
particle goes to Alice ($\rvx$). The other goes to Bob ($\rvy$).
The density matrix called $\rho$ above corresponds to a situation in which 
Bob ignores his particle. The particle is still measured passively by the
environment. Alice gets no information from the environment, so her
particle has a $50/50$ chance of being either up or down along any direction.
The density matrix called  $\rho(y)$ above corresponds to a situation in which 
instead of ignoring his particle, Bob measures it along the +Z direction and 
communicates the result to Alice. The experiment is repeated many times.
When Bob reports result $+_z$, Alice sticks her particle into bin Bob+, and
when he reports $-_z$, she sticks it into bin Bob$-$. Alice's particles
in bin Bob+ (ditto, bin Bob$-$) behave as if they were in pure state $\ket{-_z}$ (ditto, $\ket{+_z}$).
(Note that Alice's particle points opposite to Bob's. This is expected since the
initial state $\psi_{EPR}$ of the two particles has zero total spin, 
and  this quantity is conserved during the experiment.)

\EndSection


\BeginSection{Quantum Eraser}\label{sec:q-eraser}

In this section, we will consider a QB net that represents 
a situation in which one member of an EPR  pair is 
measured in a special way so as to exhibit a phenomenon 
loosely called ``quantum erasing".

Suppose $\ket{+_n}$ and  $\ket{-_n}$ are the states of spin up and down in the $+n$ direction,
where $n$ is either X or Z.
Let

\beq
\ket{+_z} = \ket{0} = 
\left [
\begin{array}{r}
1 \\
0
\end{array}
\right ]
\;,
\eeq

\beq
\ket{-_z} = \ket{1} = 
\left [
\begin{array}{r}
0 \\
1
\end{array}
\right ]
\;,
\eeq

\beq
\ket{+_x} = \ket{0_X}=
\frac{1}{\sqrt{2}}
\left [
\begin{array}{r}
1 \\
1
\end{array}
\right ]
\;,
\eeq

\beq
\ket{-_x} = \ket{1_X}=
\frac{1}{\sqrt{2}}
\left [
\begin{array}{r}
1 \\
-1
\end{array}
\right ]
\;.
\eeq
Define $U$ by

\beq
U =
\frac{1}{\sqrt{2}}
\left[
\begin{array}{rr}
1 & 1 \\
1 & -1 
\end{array}
\right]
\;.
\eeq
Note that

\beq
U\ket{+_z} = \ket{+_x}
\;,
\eeq

\beq
U\ket{-_z} = \ket{-_x}
\;.
\eeq
Also note that for $y, r \in Bool$,

\beq
\av{r | U | y} = 
\frac{1}{\sqrt{2}} (-1)^{y r}
\;.
\eeq

\begin{figure}[h]
	\begin{center}
	\epsfig{file=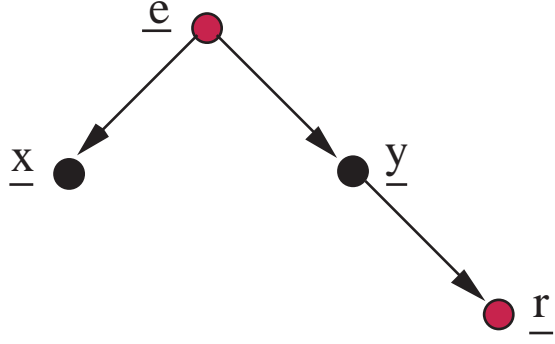}
	\caption{QB net for a quantum eraser.}
	\label{fig:eraser}
	\end{center}
\end{figure}

Consider the QB net of Fig.(\ref{fig:eraser}), where
\BeginBNetTabular
	$\rve$ & $e = (e_1, e_2) \in Bool^2$ & $\psi_{EPR}(e) = \frac{1}{\sqrt{2}}[ \delta^{e_1,e_2}_{0,1} - \delta^{e_1,e_2}_{1,0}]$ & \\
\hline
	$\rvx$ & $x\in Bool$ & $\delta(x, e_1)$ & \\
\hline
	$\rvy$ & $y\in Bool$ & $\delta(y, e_2)$ & \\
\hline
	$\rvr$ & $r\in Bool$ & $U(r  |  y) = \frac{1}{\sqrt{2}} (-1)^{y r}$ &  \\
\EndBNetTabular

Let $\qnet$ be the QB net which contains all the nodes shown 
in Fig.(\ref{fig:eraser}). Let $\qnet_0$ be the sub-net which contains only nodes $\rvx, \rve, \rvy$.

The meta density matrix $\mu_0$ of $\qnet_0$ was given 
in Section \ref{sec:EPR-pair}. Let $\rho_0 = \esum{\rve} \mu_0$. Thus,
$\rho_0$
corresponds to what we called simply $\rho$ in Section \ref{sec:EPR-pair}.

The meta density matrix $\mu$ of  $\qnet$ is

\beq
\mu = \rhometa
\;,
\eeq
where

\beq
\ketmeta = \sum_{ri} U(r  |  y) \psi_{EPR}(x, y) \ket{ \rve = (x, y), x, y, r}
\;.
\eeq
Define $\rho$ by:

\beq
\rho = \esum{\rve, \rvy} \mu
\;.
\eeq
Then

\beq
\rho = \sum_{ri} \frac{1}{4} [ (-1)^r\delta^x_0 - \delta^x_1 ]
[ (-1)^{r'}\delta^{x'}_0 - \delta^{x'}_1 ]
\ket{x, r} \bra{x', r'}
\;,
\eeq

\beq
[\bra{x,r} \rho \ket{x', r'}] = 
\frac{1}{4}
\left[
\begin{array}{r}
1\\ -1\\ -1\\ -1
\end{array}
\right ]
\left[
\begin{array}{rrrr}
1 & -1 & -1 & -1
\end{array}
\right ]
=
\frac{1}{4}
\begin{tabular}{r|rrrr}
         & {\tiny 00} & {\tiny 01} & {\tiny 10} & {\tiny 11}\\
\hline
{\tiny 00}&   1&  -1&  -1&  -1\\
{\tiny 01}&  -1&   1&   1&   1\\
{\tiny 10}&  -1&   1&   1&   1\\
{\tiny 11}&  -1&   1&   1&   1\\
\end{tabular}
\;.
\eeq
$\rho$
is a pure state so $S_\rho(\rvx, \rvr) = 0$ and $S_\rho(\rvx) = S_\rho(\rvr)$.
It is easy to show that

\beq
\tr_\rvx \rho = \frac{1}{2} \sum_r \ket{r}\bra{r}
\;,
\eeq

\beq
\tr_\rvr \rho = \frac{1}{2} \sum_x \ket{x}\bra{x}
\;,
\eeq
Thus,

\beq
\begin{tabular}{lll}
$S_\rho(\rvx) = 1,$ 			& $H_\rho(\rvx) = 1$			&	(zero coherence) \\
$S_\rho(\rvr) = 1,$			& $H_\rho(\rvr) = 1$		&	(zero coherence) \\
$S_\rho(\rvx, \rvr) = 0,$ 	& $H_\rho(\rvx, \rvr) = 2$	&	(max. coherence) \\
\\
$S_\rho(\rvx |  \rvr) = -1,$ 	& $H_\rho(\rvx |  \rvr) = 1$ &\\
$S_\rho(\rvr |  \rvx) = -1,$	& $H_\rho(\rvr |  \rvx) = 1$ &\\
$S_\rho(\rvx: \rvr) =2,$ 		& $H_\rho(\rvx: \rvr) = 0$ &\\
\end{tabular}
\;.
\eeq

Define $\rho(r)$ by

\beq
\rho(r) = 2\esum{\rve, \rvy}\bra{r} \mu \ket{r} = 2\bra{r} \rho \ket{r}
\;.
\eeq
$\rho(r)$ acts on $\hil_\rvx$. It is easy to show that

\beq
\rho(r) = \ket{0_X}\bra{0_X} \delta^r_1 + \ket{1_X}\bra{1_X} \delta^r_0 
\;.
\eeq
Thus,

\beq
\begin{tabular}{lll}
$S_{\rho(r)}(\rvx) = 0,$ 			& $H_{\rho(r)}(\rvx) = 1$			&	(max. coherence)
\end{tabular}
\;.
\label{eq:QER-ents-rho-marky}\eeq

\begin{figure}[h]
	\begin{center}
	\epsfig{file=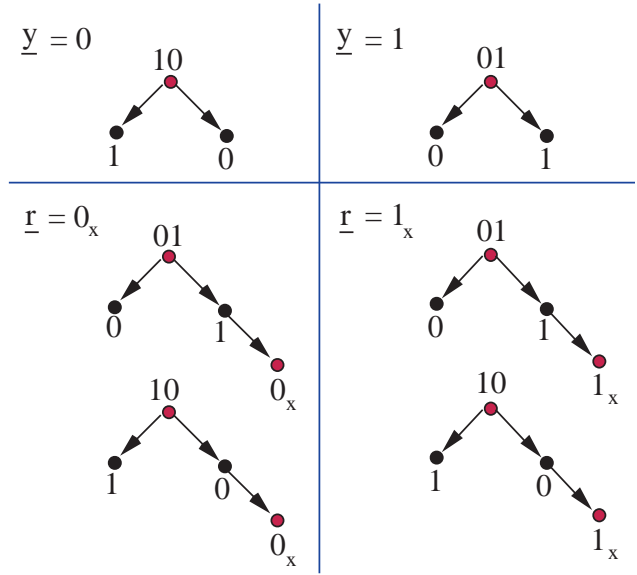, height = 3in}
	\caption{Comparison of Feynman stories for QB net Fig.(\ref{fig:pair}) representing an EPR pair and QB net Fig.(\ref{fig:eraser}) representing a quantum eraser.}
	\label{fig:eraser-stories}
	\end{center}
\end{figure}

These results can be interpreted as follows. We start with an EPR pair of particles. One 
particle goes to Alice ($\rvx$). The other goes to Bob ($\rvy, \rvr$).
Bob passes his particle through a
Stern-Gerlach magnet that separates it into its $\pm_x$ parts.
The density matrix called $\rho$ above corresponds to a situation in which 
Bob ignores his particle
after it leaves the Stern-Gerlach magnet. 
The particle is still measured passively by the
environment. Alice gets no information from the environment, so here
particle has a $50/50$ chance of being either up or down along any direction.
The density matrix called  $\rho(r)$ above corresponds to a situation in which 
instead of ignoring his particle, Bob
 measures it along the +X direction and 
communicates the result to Alice. The experiment is repeated many times.
When Bob reports result $+_x$, Alice sticks her particle into bin Bob+, and
when he reports $-_x$, she sticks it into bin Bob$-$. Alice's particles
in bin Bob+ (ditto, bin Bob$-$) behave as if they were in pure state $\ket{-_x}$ (ditto, $\ket{+_x}$).
(Note that Alice's particle points opposite to Bob's. This is expected since the
initial state $\psi_{EPR}$ of the two particles has zero total spin,
and  this quantity is conserved during the experiment.)

This is all very similar to Section \ref{sec:EPR-pair}. But note that in Section \ref{sec:EPR-pair},
Alice's particle ends in state $+_z$ (or $-_z$, depending
on the result of Bob's measurement), whereas now it ends in state $+_x$ (or $-_x$).
As shown in Fig.(\ref{fig:eraser-stories}), if the value of $\rvy$ is fixed,
then there is only one possible Feynman story. On the other hand,
if the value of $\rvr$ is fixed, there are two possible Feynman stories.
A related fact: In Section \ref{sec:EPR-pair}, Alice's particle ends in a state
characterized by the density matrix
$\ket{+_z}\bra{+_z}$ which is diagonal in the $\ket{\pm_z}$ basis, whereas now it ends
in a state  
characterized by a density matrix $\ket{+_x}\bra{+_x}$ which isn't diagonal in the $\ket{\pm_z}$ basis.

We often say that an experiment of this sort is a ``quantum eraser".
By this, we mean the following. According to Eqs.(\ref{eq:EPR-ents-rho}) and (\ref{eq:QER-ents-rho-marky})

\beq
\begin{tabular}{lll}
$S_{\rho_0}(\rvx) = 1,$ 			& $H_{\rho_0}(\rvx) = 1$			& (zero coherence) \\
\end{tabular}
\;,
\label{eq:QER-sub}\eeq

\beq
\begin{tabular}{lll}
$S_{\rho(\rvr)}(\rvx) = 0,$ 	& $H_{\rho(\rvr)}(\rvx) = 1$ 	& (max. coherence)\\
\end{tabular}
\;.
\label{eq:QER-full}\eeq
In Eq.(\ref{eq:QER-sub}), Bob ignores his particle. In Eq.(\ref{eq:QER-full}), he passes it through a Stern-Gerlach
magnet and reports the result of his measurement to Alice.
We can go from minimum coherence (Eq.(\ref{eq:QER-sub})) to the maximum coherence (Eq.(\ref{eq:QER-full}))
for node $\rvx$ simply by asking Bob to do some extra processing.
This extra processing seems to erase 
the coherence destroying mechanism. 

Note that the density matrix $\rho$ defined above acts on
$\hil_{\rvx, \rvr}$ and that 

\beq
\bra{x} \;\bra{r} \rho \ket{r} \;  \ket{x} =
\bra{r} \; \bra{x} \rho \ket{x} \; \ket{r}
\;.
\eeq
That is, the order in which we apply $\redmat{\ket{x}\bra{x}}$ and
$\redmat{\ket{r}\bra{r}}$ does not matter. 
This is often called the ``delayed choice" phenomenon.

Note that we found $H_\rho(\rvx : \rvr) = 0$ in this section,
whereas we found $H_{\rho_0}(\rvx : \rvy) = 1$ in Section \ref{sec:EPR-pair}.
That is, $\rvx$ and $\rvr$ are independent whereas 
$\rvx$ and $\rvy$ aren't. That's because
$\rvx$ and $\rvy$ must have opposite values whereas 
$\rvx$ and $\rvr$ don't have to.

\EndSection


\BeginSection{Teleportation}\label{sec:teleportation}

In this section, we will consider a QB net that represents the phenomenon known as Teleportation\cite{Tele}.

\begin{figure}[h]
	\begin{center}
	\epsfig{file=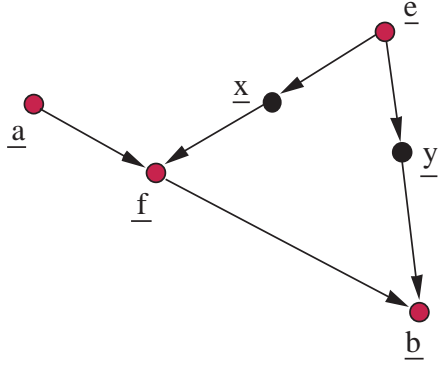, height=2.0in}
	\caption{QB net for Teleportation.}
	\label{fig:tele}
	\end{center}
\end{figure}

Consider the QB net of Fig.(\ref{fig:tele}), where
\BeginBNetTabular
	$\rve$ & $e = (e_1, e_2) \in Bool^2$ & $\psi_{EPR}(e) = \frac{1}{\sqrt{2}}[ \delta^{e_1,e_2}_{0,1} - \delta^{e_1,e_2}_{1,0}]$ & \\
\hline
	$\rvx$ & $x\in Bool$ & $\delta(x, e_1)$ & \\
\hline
	$\rvy$ & $y\in Bool$ & $\delta(y, e_2)$ & \\
\hline
	$\rva$ & $a\in Bool$ & $\alpha_a$ & $\sum_a |\alpha_a|^2= 1$\\
\hline
	$\rvf$ & $f = (f_1, f_2) \in Bool^2$ & $U(f  |  a, x)$ & $U$ specified below\\
\hline
	$\rvb$ & $b\in Bool$ & $R(b |  f, y)$ & $R$  specified below\\
\EndBNetTabular

Consider the so called ``Bell basis" vectors $\ket{\Psi(f)}$:

\beq
\ket{\Psi(f)} = \frac{1}{\sqrt{2}} [ \ket{0, f_1} + (-1)^{f_2} \ket{1, \bar{f_1}}]
\;,
\eeq
where $f\in Bool^2$, and $\bar{0} = 1$, $\bar{1} = 0$. 
$f_1$ tells us whether the two particles are in the same or different states
(different state iff $f_1=1$). 
$f_2$ tells us the sign between the two kets being summed
(minus sign iff $f_2=1$).
For example,

\beq
\ket{\Psi(1, 1)} = \frac{1}{\sqrt{2}} ( \ket{0, 1} - \ket{1, 0})
\;.
\eeq
The state $\psi_{EPR}(e)$ given above equals $\av{e | \Psi(1, 1) }$.

We define the matrix $U$ mentioned above by

\beq
U(f  |  a,x) = \av{\Psi(f)  |  a, x} = 
\frac{1}{\sqrt{2}} [ \delta^{a,x}_{0,f_1} + (-1)^{f_2} \delta^{a,x}_{1, \bar{f_1}} ]
\;,
\eeq

\beq
[ U(f  |  a,x) ] 
=
\frac{1}{\sqrt{2}}
\begin{tabular}{r|rrrr}
         & {\tiny 00} & {\tiny 01} & {\tiny 10} & {\tiny 11}\\
\hline
{\tiny 00}&   1&   0&   0&   1\\
{\tiny 01}&   1&   0&   0&  -1\\
{\tiny 10}&   0&   1&   1&   0\\
{\tiny 11}&   0&   1&  -1&   0\\
\end{tabular}
\;.
\eeq
The columns of $U$ are clearly orthonormal so $U$ is a unitary matrix.

The matrix $R$ mentioned above can be defined in terms of $U$ by

\beq
R(b  |  f, y) = U( f  |  b, \bar{y}) (-1)^{\bar{y}} (-1)^{f_1 f_2} \sqrt{2}
\;.
\eeq
Our reasons for defining $R$ in this way will become clear as we go on.
Note that

\beq
\sum_b | R(b  |  f, y)|^2 =1
\;,
\eeq
as required by the definition of QB nets.

It is convenient to define a function $K(\cdot)$ by

\beq
K(x, y, a, f, b) =
R(b  |  f,y) U( f  |  a, x) \psi_{EPR}(x, y)
\;.
\eeq
Substituting explicit expressions for
$R, U$ and $\psi_{EPR}$ into the last equation yields

\beq
K(x, y, a, f, b) = \frac{ (-1)^{f_1 f_2}}{2}\;
\delta^{a, x}_{b, \bar{y}} ( \delta^{a,x}_{0, f_1} + \delta^{a,x}_{1, \bar{f_1}})
\;.
\eeq
From this expression for $K(\cdot)$, it follows that

\begin{subequations}
\label{eq:TE-K-ids}
\beq
\sum_{x, y} K = \frac{(-1)^{f_1 f_2}}{2} \delta^a_b
,\;\;\;
\sum_{x, y, f} K = \delta^a_b
\;,
\eeq

\beq
\sum_{x, y} |K|^2 = \frac{1}{4} \delta^a_b
,\;\;\;
\sum_{x, y, f} |K|^2 = \delta^a_b
\;.
\eeq
\end{subequations}

Define the following kets:

\begin{subequations}
\label{eq:TE-kets}
\beq
\ket{\psi_{in}} = \sum_a \alpha_a \ket{\rva =a }
\;,
\eeq

\beq
\ket{\psi'_{in}} = \sum_a \alpha_a \ket{\rvb =a }
\;,
\eeq

\beq
\ket{\psi_{out}} = \sum_{x.} A(x.) \ket{(x.)_\zex }
=
\sum_{all} K(x, y, a, f, b)\alpha_a \ket{b}
\;,
\eeq

\beq
\ket{\psi_{out}(f)} = 2 \sum_{all/f} K(x, y, a, f, b)\alpha_a \ket{b}
\;.
\eeq
\end{subequations}
Note that we don't sum over $f$ in the equation for $\ket{\psi_{out}(f)}$.
It follows by Eqs.(\ref{eq:TE-K-ids}) that the kets of Eqs.(\ref{eq:TE-kets}) have unit magnitude and that

\beq
\ket{\psi_{out}(f)} = (-1)^{f_1 f_2}\ket{\psi'_{in}} 
\;,
\label{eq:TE-tele}\eeq

\beq
\ket{\psi_{out}} = \ket{\psi'_{in}} 
\;.
\label{eq:TE-pseudo-tele}\eeq
Because of Eq.(\ref{eq:TE-tele}), one says that the QB net of Fig.(\ref{fig:tele})
``teleports"  a quantum state from node $\rva$ to node $\rvb$.
Without knowing the state $\ket{\psi_{in}}$, Alice at $\rvf$ measures the joint state delivered to her
by $\rva$ and $\rvx$. She obtains result $f$ which she
sends by classical means to Bob at $\rvb$. 
Bob can choose to allow any value of $f$, or 
he can ignore those repetitions of the experiment in which 
$f$ does not equal a particular value, say $(0,1)$.
In either case, the state
$\ket{\psi_{out}(f)}$ emerging from Bob's lab $\rvb$ is 
equal to $\pm\ket{\psi'_{in}}$.
Note that
according to Eq.(\ref{eq:TE-pseudo-tele}), 
even if Alice does not measure $\rvf$, and instead she sends a quantum message to Bob, 
$\ket{\psi_{out}}$ equals $\ket{\psi'_{in}}$. However, this is not ``true" teleportation.
In ``true" teleportation, we allow Alice to receive quantum messages but not to send them.

The meta density matrix $\mu$ for the net of Fig.(\ref{fig:tele}) is

\beq
\mu = \rhometa
\;,
\eeq
where

\beq
\ketmeta = \sum_{all} K(x,y,a,f,b) 
\alpha_a
\ket{ \rve = (x, y), x, y, a , f, b}
\;.
\eeq
Note that by Eqs.(\ref{eq:TE-K-ids}), $\ketmeta$ has unit magnitude.

Define the reduced matrix $\sigma$ by

\beq
\sigma = \esum{\rve, \rvx, \rvy} (\mu)
\;.
\eeq
It is easy to show that

\beq
\sigma = 
\ket{\phi_{\rva,\rvb}} \bra{ \phi_{\rva,\rvb}}
\;\;\;
\ket{\phi_{\rvf}} \bra{ \phi_{\rvf}}
\;,
\eeq
where

\beq
\ket{ \phi_{\rva,\rvb}} = \sum_a \alpha_a \ket{\rva=a,\rvb=a}
\;,
\eeq

\beq
\ket{ \phi_{\rvf}} = \sum_f \frac{(-1)^{f_1 f_2}}{2} \;\ket{f}
\;.
\eeq

Define 

\beq
H^{in} = -\sum_{a\in Bool} |\alpha_a|^2 \log_2 (|\alpha_a|^2)
\;.
\eeq
Next we will calculate classical and
quantum entropies for various possible density matrices $\rho$:

\MyCases{(a)$\rho = \trace{\rvb} \sigma$}

Then 

\beq
\rho = \left( \sum_a |\alpha_a|^2 \ket{a}\bra{a} \right)
\ket{\phi_\rvf} \bra{\phi_\rvf}
\;.
\label{eq:TE-trace-b}\eeq
It is easy to show from Eq.(\ref{eq:TE-trace-b}) that

\beq
\begin{tabular}{lll}
$S_\rho(\rva) = H^{in},$ 			& $H_\rho(\rva) = H^{in}$			&	(zero coherence) \\
$S_\rho(\rvf) = 0,$					& $H_\rho(\rvf) = 2$				&	(max. coherence) \\
$S_\rho(\rva, \rvf) = H^{in},$ 		& $H_\rho(\rva, \rvf) = H^{in} + 2$	&\\
\\
$S_\rho(\rva |  \rvf) = H^{in},$ 	& $H_\rho(\rva |  \rvf) = H^{in}$ 	&\\
$S_\rho(\rvf |  \rva) = 0,$		& $H_\rho(\rvf |  \rva) = 2$ 		&\\
$S_\rho(\rva: \rvf) =0,$ 			& $H_\rho(\rva: \rvf) = 0$ 			&\\
\end{tabular}
\;.
\eeq

\MyCases{(b)$\rho = {\cal N}\bra{f}\sigma\ket{f}$}

Then 

\beq
\rho = 
\ket{\phi_{\rva, \rvb}} 
\bra{\phi_{\rva, \rvb}}
\;.
\label{eq:TE-trace-f}\eeq
Note that we get the same density matrix
if we reduce $\sigma$ by projecting, tracing or e-summing over node $\rvf$:

\beq
{\cal N}\bra{f}\sigma\ket{f} = \trace{\rvf}\sigma = 
\esum{\rvf} \sigma
\;.
\eeq
It is easy to show from Eq.(\ref{eq:TE-trace-f}) that

\beq
\begin{tabular}{lll}
$S_\rho(\rva) = H^{in},$ 			& $H_\rho(\rva) = H^{in}$			&	(zero coherence) \\
$S_\rho(\rvb) = H^{in},$			& $H_\rho(\rvb) = H^{in}$			&	(zero coherence) \\
$S_\rho(\rva, \rvb) = 0,$ 			& $H_\rho(\rva, \rvb) = H^{in}$		&\\
\\
$S_\rho(\rva |  \rvb) = -H^{in},$ 	& $H_\rho(\rva |  \rvb) = 0$ 		&\\
$S_\rho(\rvb |  \rva) = -H^{in},$	& $H_\rho(\rvb |  \rva) = 0$ 		&\\
$S_\rho(\rva: \rvb) =2H^{in},$ 		& $H_\rho(\rva: \rvb) = H^{in}$ 	& {\tiny transmitted info: quantum  = 2 classical}\\
\end{tabular}
\;.
\eeq

\EndSection


\BeginSection{Qubit Bouncing (a.k.a. Dense Coding)}\label{sec:qbit-bouncing}

Ref.\cite{QubitBouncing} was the first to discuss a phenomenon that we will call qubit bouncing. 
Qubit bouncing is often called ``quantum super dense coding".
In this section, we will consider a QB net that represents qubit bouncing.

\begin{figure}[h]
	\begin{center}
	\epsfig{file=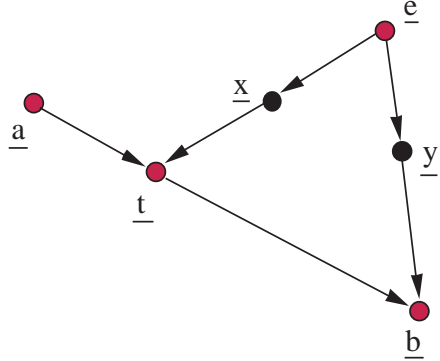, height=2.0in}
	\caption{QB net for Qubit Bouncing.}
	\label{fig:bounce}
	\end{center}
\end{figure}

Consider the QB net of Fig.(\ref{fig:bounce}), where
\BeginBNetTabular
	$\rve$ & $e = (e_1, e_2) \in Bool^2$ & $\psi_{EPR}(e) = \frac{1}{\sqrt{2}}[ \delta^{e_1,e_2}_{0,1} - \delta^{e_1,e_2}_{1,0}]$ & \\
\hline
	$\rvx$ & $x\in Bool$ & $\delta(x, e_1)$ & \\
\hline
	$\rvy$ & $y\in Bool$ & $\delta(y, e_2)$ & \\
\hline
	$\rva$ & $a=(a_1, a_2)\in Bool^2$ & $\alpha_a$ & $\sum_a |\alpha_a|^2= 1$\\
\hline
	$\rvt$ & $t \in Bool$ & $R(t  |  a, x)$ & $R$ specified below\\
\hline
	$\rvb$ & $b=(b_1, b_2)\in Bool^2$ & $U(b |  t, y)$ & $U$ specified below\\
\EndBNetTabular

The matrix $U$ in this section is identical to its namesake in the Teleportation section:

\beq
U(b  |  t,y) = 
\frac{1}{\sqrt{2}} ( \delta^{t, y}_{0,b_1} + (-1)^{b_2} \delta^{t, y}_{1, \bar{b_1}} )
\;.
\eeq

The matrix $R$ can be defined in terms of $U$ by

\beq
R(t  |  a, x) = U( a  |  t, \bar{x}) (-1)^{x} \sqrt{2}
\;.
\eeq
Our reasons for defining $R$ in this way will become clear as we go on.
Note that

\beq
\sum_t | R(t  |  a, x)|^2 =1
\;,
\eeq
as required by the definition of QB nets.

It is convenient to define a function $K(\cdot)$ by

\beq
K(x, y, a, t, b) =
U( b  |  t, y) R(t  |  a,x) \psi_{EPR}(x, y)
\;.
\eeq
Substituting explicit expressions for
$R, U$ and $\psi_{EPR}$ into the last equation yields

\beq
K(x, y, a, t, b) = \frac{ 1}{2}\;
\delta^{a_1, y}_{b_1, \bar{x}} [ \delta^{t,y}_{0, a_1} + (-1)^{a_2 + b_2} \delta^{t,y}_{1, \bar{a}_1}]
\;.
\eeq
From this expression for $K(\cdot)$, it follows that

\begin{subequations}
\label{eq:QB-K-ids}
\beq
\sum_{x, y} K = \frac{1}{2} \delta^{a_1}_{b_1}
[\delta^t_0 + (-1)^{a_2 + b_2} \delta^t_1]
,\;\;\;
\sum_{x, y, t} K = \delta^a_b
\;,
\eeq

\beq
\sum_{x, y} |K|^2 = \frac{1}{4} \delta^{a_1}_{b_1}
,\;\;\;
\sum_{x, y, t, b} |K|^2 = 1
\;.
\eeq
\end{subequations}

Define the following kets:

\begin{subequations}
\label{eq:QB-kets}
\beq
\ket{\psi_{in}} = \sum_a \alpha_a \ket{\rva =a }
\;,
\eeq

\beq
\ket{\psi'_{in}} = \sum_a \alpha_a \ket{\rvb =a }
\;,
\eeq

\beq
\ket{\psi_{out}} = \sum_{x.} A(x.) \ket{(x.)_\zex }
=
\sum_{all} K(x, y, a, t, b)\alpha_a \ket{b}
\;,
\eeq
\end{subequations}
It follows by Eqs.(\ref{eq:QB-K-ids}) that the kets of Eqs.(\ref{eq:QB-kets}) have unit magnitude and that

\beq
\ket{\psi_{out}} = \ket{\psi'_{in}} 
\;.
\eeq

The meta density matrix $\mu$ for the net of Fig.(\ref{fig:bounce}) is

\beq
\mu = \rhometa
\;,
\eeq
where

\beq
\ketmeta = \sum_{all} K(x,y,a,t,b) 
\alpha_a
\ket{ \rve = (x, y), x, y, a , t, b}
\;.
\eeq
Note that by Eqs.(\ref{eq:QB-K-ids}), $\ketmeta$ has unit magnitude.

Define the reduced matrix $\sigma$ by

\beq
\sigma = \esum{\rve, \rvx, \rvy} (\mu)
\;.
\eeq
It is easy to show that
\beq
\sigma = 
\ket{\phi_{\rva,\rvt, \rvb}} \bra{ \phi_{\rva,\rvt, \rvb}}
\;,
\eeq
where

\beq
\ket{ \phi_{\rva,\rvt, \rvb}} = \sum_{all}
\frac{1}{2} \delta^{a_1}_{b_1}
[\delta^t_0 + (-1)^{a_2 + b_2} \delta^t_1]
\alpha_a \ket{a, t, b}
\;,
\eeq

Define 

\beq
H^{in} = -\sum_{a\in Bool^2} |\alpha_a|^2 \log_2 (|\alpha_a|^2)
\;,
\eeq

\beq
w_{a_1} = \sum_{a_2} |\alpha_{a_1 a_2}|^2
\;,
\eeq

\beq
H^{in}_1 =  -\sum_{a_1\in Bool} w_{a_1} \log_2 (w_{a_1})
\;.
\eeq

Next we will calculate classical and
quantum entropies for various possible density matrices $\rho$:

\MyCases{(a) $\rho = \trace{\rvb} \sigma$}

Then

\begin{subequations}
\label{eq:QB-trace-b}
\beq
\rho = \sum_{a_1, t} \left( w_{a_1, t} \ket{a_1, t} \bra{a_1, t} \right)\rho_{a_1,t}
\;,
\eeq
where

\beq
w_{a_1, t} = \frac{1}{2} w_{a_1}
\;,
\eeq

\beq
\rho_{a_1, t} = \ket{\phi_{\rva_2}(a_1, t)}\bra{\phi_{\rva_2}(a_1, t)}
\;,
\eeq
where

\beq
\ket{\phi_{\rva_2}(a_1, t)} =
\frac{1}{\sqrt{w_{a_1}}} \sum_{a_2} 
[\delta^t_0 + (-1)^{a_2} \delta^t_1]\alpha_{a_1 a_2}
\ket{a_2}
\;.
\eeq
\end{subequations}
It is easy to show from Eqs.(\ref{eq:QB-trace-b}) that
\beq
\begin{tabular}{lll}
$S_\rho(\rva) = H^{in},$ 						& $H_\rho(\rva) = H^{in}$			&	(zero coherence) \\
$S_\rho(\rvt) = 1,$								& $H_\rho(\rvt) = 1$				&	(zero coherence) \\
$S_\rho(\rva, \rvt) = 1+ H^{in}_1,$ 			& $H_\rho(\rva, \rvt) = 1+ H^{in}$	&\\
\\
$S_\rho(\rva |  \rvt) = H^{in}_1,$ 				& $H_\rho(\rva |  \rvt) = H^{in}$ 	&\\
$S_\rho(\rvt |  \rva) = 1 + H^{in}_1 - H^{in},$	& $H_\rho(\rvt |  \rva) = 1$ 		&\\
$S_\rho(\rva: \rvt) =H^{in} - H^{in}_1,$ 		& $H_\rho(\rva: \rvt) = 0$ 			&\\
\end{tabular}
\;.
\eeq

\MyCases{(b) $\rho = \esum{\rvt} \sigma$} 

Then 

\begin{subequations}
\label{eq:QB-trace-t}
\beq
\rho = \ket{\phi_{\rva, \rvb}} \bra{\phi_{\rva, \rvb}}
\eeq
where

\beq
\ket{\phi_{\rva, \rvb}} = \sum_a \alpha_a \ket{\rva=a,\rvb=a}
\;.
\eeq
\end{subequations}
It is easy to show from Eqs.(\ref{eq:QB-trace-t}) that

\beq
\begin{tabular}{lll}
$S_\rho(\rva) = H^{in},$ 			& $H_\rho(\rva) = H^{in}$			&	(zero coherence) \\
$S_\rho(\rvb) = H^{in},$			& $H_\rho(\rvb) = H^{in}$			&	(zero coherence) \\
$S_\rho(\rva, \rvb) = 0,$ 			& $H_\rho(\rva, \rvb) = H^{in}$		&\\
\\
$S_\rho(\rva |  \rvb) = -H^{in},$ 	& $H_\rho(\rva |  \rvb) = 0$ 		&\\
$S_\rho(\rvb |  \rva) = -H^{in},$	& $H_\rho(\rvb |  \rva) = 0$ 		&\\
$S_\rho(\rva: \rvb) =2H^{in},$ 		& $H_\rho(\rva: \rvb) = H^{in}$ 	& {\tiny transmitted info: quantum  = 2 classical}\\
\end{tabular}
\;.
\eeq

\EndSection

\begin{appendix}


\BeginSection{Review of Classical and Quantum \\  Bayesian Nets}\label{app:bnet-review}

In this Appendix, we give a brief review of Classical Bayesian (CB) and Quantum Bayesian (QB) nets. For more information,
see Ref.\cite{Tucci95}.

First, we will state those properties which 
CB and QB nets have in common. 

We call a {\it graph} (or a diagram ) a collection of nodes with arrows connecting some
pairs of these nodes. The arrows of the graph must satisfy certain
constraints that will be specified below. We call a {\it labelled graph}
a graph whose nodes are
labelled.  A
{\it CB net} (ditto, a {\it QB net}) consists of  two parts: a  labelled graph with each node
labelled by a random variable,  and a collection of node matrices, one 
matrix for each node. These two parts must satisfy 
certain constraints that will be specified below.

An {\it internal arrow} is an arrow that
has  a starting (source) node and a different ending (destination) one.
We will use only internal arrows.
We  define two
types of nodes:  an {\it internal node} is a node that has
one or more internal arrows leaving it, and an {\it external node} is a node that has no
internal arrows leaving it. It is also common to
use the terms {\it root node} or {\it prior probability node} for a node
which has no incoming arrows (if any arrows touch it, they are outgoing ones). 

We restrict our attention to {\it acyclic} graphs; that is, graphs that do not contain cycles.
(A {\it cycle} is a closed path of arrows with the arrows all pointing in the same sense.)

 We assign a
 random variable to each node of a CB net. 
 Suppose the random
variables assigned to the $N$ nodes are $\rvx_1,\rvx_2,\cdots,\rvx_N$. For each $j\in \zn$, 
the random variable $\rvx_j$ will be assumed to take on
values  within a finite set $S_{\rvx_j}$ called
{\it the set of possible states of} $\rvx_j$. 

If $\Gamma=\{k_1,k_2,\cdots, k_{|\Gamma|}\}\subset \zn$,
and $k_1< k_2 < \cdots < k_{|\Gamma|}$,
define $(x.)_\Gamma=(x_{k_1},x_{k_2},\cdots,x_{k_{|\Gamma|}})$ and 
$(\rvx.)_\Gamma=(\rvx_{k_1},\rvx_{k_2},\cdots,\rvx_{k_{|\Gamma|}})$.
Sometimes, we also abbreviate $(x.)_{\zn}$ 
(i.e., the vector that includes all the possible $x_j$ components) by just
$x.$, and $(\rvx.)_{\zn}$ by just $\rvx.\;$. We often refer to 
$\rvX = (\rvx.)_\Gamma$ as a {\it node collection}. We say 
$\rvX$ is empty if $|\Gamma|= 0$. If $|\Gamma|=1$,
we say it is a {\it single-node} node collection, and if
$|\Gamma|>1$, we say it is a {\it compound} node collection.
Given two node collections  $\rvX_1 = (\rvx.)_{\Gamma_1}$ and
$\rvX_2 = (\rvx.)_{\Gamma_2}$, we say that $\rvX_1$ and $\rvX_2$ are disjoint
(ditto, $\rvX_1$ is a subset of $\rvX_2$), if
$\Gamma_1$ and $\Gamma_2$ are disjoint (ditto, $\Gamma_1\subset\Gamma_2$).

Let $\zex$ be the set of all $j\in \zn$ such that $\rvx_j$ is
an external node,  and  let $\zin$ be the set of all $j\in \zn$ such
that 
 $\rvx_j$  is an internal node. Clearly, $\zex$ and $\zin$ are disjoint
and their union is $\zn$. 

Each possible value $x.$ of 
$\rvx.$ defines a different {\it net story}.
 For any net story $x.$,
we call $(x.)_\zin$ the {\it internal state of the story} and 
$(x.)_\zex$ its {\it external state}.

Define $\Gamma_j$ to be the set of all $k$ such that 
an arrow labelled $x_k$ (i.e., an arrow whose source node is $\rvx_k$) enters node 
$\rvx_j$. 

Next, we will state those properties which 
are different in CB and QB nets.
 
\MyCases{(a) Classical Bayesian Net}

 For each net story $x.$ of a CB net, we 
assign a non-negative number $P_j[x_j|(x.)_{\Gamma_j}]$ to each node $\rvx_j$.
 We call
$P_j[x_j|(x.)_{\Gamma_j}]$  the {\it probability of node} $\rvx_j$
{\it within net story} $x.$. 
The function $P_j$ with values $P_j[x_j|(x.)_{\Gamma_j}]$ determines a matrix that we 
call the  {\it node matrix of node} $\rvx_j$.
 $x_j$ is the matrix's  {\it row index} and   $(x.)_{\Gamma_j}$ is
its {\it column index}. We require that the values $P_j[x_j|(x.)_{\Gamma_j}]$
be conditional probabilities; i.e., that they satisfy:

\beq
P_j[x_j | (x.)_{\Gamma_j}] \geq 0
\;,
\label{eq:RV-geq-0}\eeq

\beq
\sum_{x_j} P_j[x_j | (x.)_{\Gamma_j}] = 1
\;,
\label{eq:RV-norm-pj}\eeq
where the sum in Eq.(\ref{eq:RV-norm-pj}) is over all the states 
that the random variable $\rvx_j$ can assume, and where
Eqs.(\ref{eq:RV-geq-0}) and (\ref{eq:RV-norm-pj}) must be satisfied for all $j\in\zn$
and for all possible values of the vector $(\rvx.)_{\Gamma_j}$ of random variables.
The left-hand side of Eq.(\ref{eq:RV-norm-pj}) is just the sum over  the entries
of a column of the node matrix.

The {\it probability of net story} $x.$,
call it $P(x.)$,  is  defined to 
be the product of all the node probabilities 
$P_j[x_j|(x.)_{\Gamma_j}]$ for  $j\in \zn$. Thus,

\beq
P(x.)=\prod_{j\in \zn}
P_j[x_j|(x.)_{\Gamma_j}]
\;.
\label{eq:RV-def-joint}\eeq
We require $P(x.)$ to satisfy:

\beq
\sum_{x.} P(x.) = 1
\;.
\label{eq:RV-norm-p}\eeq 

Call a {\it CB  pre-net} a labelled graph and an
accompanying set of node matrices that satisfy Eqs.(\ref{eq:RV-geq-0}), (\ref{eq:RV-norm-pj}) and
(\ref{eq:RV-def-joint}), but don't necessarily  satisfy the overall
normalization condition Eq.(\ref{eq:RV-norm-p}). It can be shown that all acyclic  CB pre-nets
satisfy Eq.(\ref{eq:RV-norm-p}). If one considers only acyclic
graphs as we do in this paper, then there is no 
difference between CB
nets  and CB pre-nets. 

\MyCases{(b) Quantum Bayesian Net}

For each net story $x.$ of a QB net, we may assign a
a complex number $A_j[x_j|(x.)_{\Gamma_j}]$ to each node $\rvx_j$.
 We call
$A_j[x_j|(x.)_{\Gamma_j}]$  the {\it amplitude of node} $\rvx_j$
{\it within net story} $x.$. 
The function $A_j$ with values $A_j[x_j|(x.)_{\Gamma_j}]$ determines a matrix that we 
call the  {\it node matrix of node} $\rvx_j$.
 $x_j$ is the matrix's  {\it row index} and   $(x.)_{\Gamma_j}$ is
its {\it column index}.  We require that the quantities $A_j[x_j|(x.)_{\Gamma_j}]$
be probability amplitudes that satisfy:

\beq
\sum_{x_j} \left|A_j[x_j | (x.)_{\Gamma_j}]\right|^2 = 1
\;,
\label{eq:RV-norm-aj}\eeq
where the sum in Eq.(\ref{eq:RV-norm-aj}) is over all the states 
that the random variable $\rvx_j$ can assume, and where
Eq. (\ref{eq:RV-norm-aj}) must be satisfied for all $j\in\zn$
and for all possible values of the vector $(\rvx.)_{\Gamma_j}$ of random variables.

The {\it amplitude of net story} $x.$,
call it $A(x.)$,  is  defined to 
be the product of all the node amplitudes 
$A_j[x_j|(x.)_{\Gamma_j}]$ for  $j\in \zn$. Thus,

\beq
A(x.)=\prod_{j\in \zn}
A_j[x_j|(x.)_{\Gamma_j}]
\;.
\label{eq:RV-def-a}\eeq
We require $A(x.)$ to satisfy:

\beq
\sum_{(x.)_\zex} \left| \sum_{(x.)_\zin}  A(x.) \right |^2 =1
\;
\label{eq:RV-io-norm-a}\eeq
and

\beq
\sum_{x.} |A(x.)|^2 = 1
\;.
\label{eq:RV-norm-a}\eeq

Note that as a consequence of Eqs.(\ref{eq:RV-norm-aj}) and (\ref{eq:RV-norm-a}), given any QB
net, one can construct a special CB net by replacing at each node the value
$A[x_j|(x.)_{\Gamma_j}]$ by its magnitude squared. 
We call this special CB
net  the {\it parent CB net}
 of the  QB net from which it was
constructed. We call it so because, given  a parent CB net, one
can  replace the  value of each node by its square root times a phase
factor. For a different choice of phase factors, one  generates a
different QB net. Thus, a parent CB net may be used to generate a
whole family of QB nets.

A {\it QB pre-net} is a labelled graph and an accompanying set of
node matrices that satisfy Eqs.(\ref{eq:RV-norm-aj}), (\ref{eq:RV-def-a}) and (\ref{eq:RV-io-norm-a}),  but  don't
necessarily satisfy Eq.(\ref{eq:RV-norm-a}). A QB pre-net that is acyclic satisfies
Eq.(\ref{eq:RV-norm-a}), because its parent CB pre-net is acyclic and this implies that
Eq.(\ref{eq:RV-norm-a}) is satisfied.  If one considers only acyclic
graphs  as we do in this paper, then there is no 
difference between QB
nets  and QB pre-nets. One can check that all the examples of QB nets 
considered in this paper satisfy
Eq.(\ref{eq:RV-norm-a}). Eq.(\ref{eq:RV-norm-a}) is true iff
the meta state $\ketmeta$ defined by Eq.(\ref{eq:ADM-ketmeta}) has unit magnitude.

\EndSection

\end{appendix}


\EndSection


\begin{thebibliography}{99}
\bibitem{pom}
It is also called a POVM, which stands for 
((Positive Operator) Valued) Measure.
The reason for the long name is as follows.

In classical probability, one speaks of an event space $\Omega$
and a function $\mu :\Omega \rarrow ({\rm Non-negative Reals})$ 
called  a real-valued measure.
A random variable $\rvb$ on $\Omega$ is a function
$\rvb: \Omega \rarrow S_\rvb$, where $S_\rvb$ is the set of
values that $\rvb$ may assume. $P(\rvb=b)$ is defined by 
	$$ P(\rvb=b) = \mu \left( \{\omega\in \Omega \mid \rvb(\omega) = b \} \right).$$

In quantum mechanics, 
one speaks of an event space $\Omega$,
a Hilbert space $\hil$,
a density matrix $\rho$ acting on $\hil$,
and a function $\mu_{op}: \Omega \rarrow ({\rm Non-negative Operators Acting On}\; \hil)$ 
called an operator-valued measure.
A random variable $\rvb$ is still 
a function $\rvb: \Omega \rarrow S_\rvb$. 
For each $b\in S_\rvb$, one defines 
an operator $F_b$ acting on $\hil$ by 
	$$F_b = \mu_{op}\left( \{\omega\in \Omega \mid \rvb(\omega) = b \} \right).$$
Then $P(\rvb=b)$ is defined by 
	$$P(\rvb = b) = \tr (F_b \rho).$$
It's really $\mu_{op}$ that is a POM, but since the set $\pom$
partly specifies $\mu_{op}$, we call this set a  POM too.
For more information about POMs, see \cite{CavesRev} and references therein.
\end{thebibliography}
\end{document}